\documentclass[aps,prb,twocolumn,groupedaddress,showpacs,floatfix]{revtex4-1}
\pdfoutput=1

\usepackage{graphicx}
\usepackage{dcolumn}
\usepackage{bm}
\usepackage{amsmath}
\DeclareMathOperator{\Tr}{Tr}
\DeclareMathOperator{\sgn}{sgn}

\begin{document}

\title{Graphene flakes with defective edge terminations: Universal and topological aspects,
and one-dimensional quantum behavior}

\author{Igor Romanovsky}
\email{Igor.Romanovsky@physics.gatech.edu}
\author{Constantine Yannouleas}
\email{Constantine.Yannouleas@physics.gatech.edu}
\author{Uzi Landman}
\email{Uzi.Landman@physics.gatech.edu}

\affiliation{School of Physics, Georgia Institute of Technology,
             Atlanta, Georgia 30332-0430}

\date{11 October 2012}

\begin{abstract}
Systematic tight-binding investigations of the electronic spectra (as a function of the magnetic field) 
are presented for trigonal graphene nanoflakes with reconstructed zigzag edges, where a succession
of pentagons and heptagons, that is 5-7 defects, replaces the hexagons at the zigzag edge. For nanoflakes 
with such reczag defective edges, emphasis is placed on topological aspects and connections underlying the 
patterns dominating these spectra. The electronic spectra of trigonal graphene nanoflakes with  reczag edge 
terminations exhibit certain unique features, in addition to those that are well known to appear for 
graphene dots with zigzag edge termination. These unique features include breaking of the particle-hole 
symmetry, and they are associated with nonlinear dispersion of the energy as a function of momentum, which 
may be interpreted as nonrelativistic behavior. The general topological features shared with the zigzag 
flakes include the appearance of energy gaps at zero and low magnetic fields due to finite size, the 
formation of relativistic Landau levels at high magnetic fields, and the presence between the Landau levels 
of edge states (the socalled Halperin states) associated with the integer quantum Hall effect. 
Topological regimes, unique to the reczag nanoflakes, appear within a stripe of negative energies 
$\varepsilon_b < \varepsilon < 0$, and along a separate feature forming a constant-energy line outside this 
stripe. The $\varepsilon_b$ lower bound specifying the energy stripe is independent of size.

Prominent among the patterns within the $ \varepsilon_b < \varepsilon < 0$ 
energy stripe is the formation of three-member braid
bands, similar to those present in the spectra of narrow graphene {\it nanorings\/}; they are associated 
with Aharonov-Bohm-type oscillations, i.e., the reczag edges along the three sides of the triangle behave
like a nanoring (with the corners acting as scatterers) enclosing the magnetic flux through the entire area
of the graphene flake. Another prominent feature within the $ \varepsilon_b < \varepsilon < 0$ 
energy stripe is a subregion of 
Halperin-type edge states of enhanced density immediately below the zero-Landau level. Furthermore, there 
are features resulting from localization of the Dirac quasiparticles at the corners of the polygonal flake.

A main finding concerns the limited applicability of the continuous Dirac-Weyl equation, since the latter
does not reproduce the special reczag features. Due to this discrepancy between the tight-binding and 
continuum descriptions, one is led to the conclusion that the linearized Dirac-Weyl equation fails to 
capture essential nonlinear physics resulting from the introduction of a multiple topological defect in the 
honeycomb graphene lattice.

\end{abstract}


\maketitle

\section{Introduction}
\label{secint}

\subsection{Edge terminations and their nanoelectronics potential}
\label{secnano}

Graphene is a single-layer honeycomb lattice of carbon atoms and exhibits
novel behavior due to the relativistic-like character of quasiparticle (particle-hole) excitations near 
the Fermi level (the Dirac neutrality point).\cite{geim04,geim05} In addition to the intrinsic interest in
this material, the potential of graphene for nanoelectronics applications has generated
considerable amount of research regarding the physics governing the Dirac electrons
in graphene nanostructures. Initially graphene nanoribbons attracted most of the
attention; see, e.g., Refs.\ \onlinecite{dres96,waka99,cai10}. However, in the last couple 
of years the focus is being shifted to studying zero-dimensional stuctures like graphene 
quantum dots and graphene quantum voids (see, e.g., Refs.\ \onlinecite{wuns08,roma09,wurm09,libi09,
zhan08,pala07,ezaw08,pota10,yan10,yann10,roma11}), as well as graphene nanorings (see, e.g., 
Refs.\ \onlinecite{rech07,baha09,roma12}).

In addition to the novelty of the relativistic nature of the trapped
quasiparticles, the honeycomb lattice of graphene provides for a variety of edge
terminations (see below), which have no parallel in the case of semiconductor nanosystems. 
More importantly, it is now understood \cite{dres11} that the electronic properties
of graphene nanostructures are drastically influenced by the character of the edge termination.

The physical graphene edges develop along the crystallographic axes of the
honeycomb lattice, and they may exhibit two distinct types of terminations: zigzag
or armchair. One-type edges may intersect at angles of $60^\circ$ or $120^\circ$,
yielding graphene flakes and voids with regular trigonal or hexagonal shapes. Square
graphene dots can also be envisioned, but they have edges of a mixed zigzag and
armchair character. Ring-like trigonal, hexagonal, and square-like graphene
structures are also the focus of intensive theoretical studies.

The theoretical advances regarding the properties of graphene edges have in turn
motivated considerable experimental efforts aiming at producing graphene edges with
a high-degree of purity with respect to the edge termination (zigzag or armchair),
and remarkable successes have been already reported; see, e.g., Refs.\ 
\onlinecite{dres11,jia09,krau10,neme10,yang10,shi11,lu11,begl11}.

While the zigzag and armchair edges were known for some time from the theoretical
studies on graphene nanoribbons, the recent consideration (anticipated theoretically
and confirmed through observation) of yet another physical edge, formed through
reconstruction of the zigzag edge, has added a new dimension to the research on the
electronic properties of graphene nanostructures.\cite{kosk08,kosk09,rodr11,osta11}
Indeed, this reconstructed edge, which is usually called reczag and consists of a succession 
of pentagons and heptagons (5$-$7 defect) according to the Stone-Wales-defects prescription, has 
the potential to yield new distinctive features in the electronic structure of graphene
nanostructures, whether these nanostructures are graphene flakes, voids, or graphene rings.
The reczag edge belongs to a general class of defective formations in graphene: a related 
defective formation is the alternation of pentagons and octagons (5$-$8$-$5 defect), which has also
been observed experimentally in the last couple of years and which is expected to behave like a 
``quantum wire'' within the graphene sheet. \cite{dres11,lahi10}  

\subsection{Topological aspects: Coexistence of quantum-wire, ideal-ring, and
quantum-dot singly-connected-geometry behavior}
\label{sectopo}

Experimentally, two-dimensional semiconductor quantum dots (SQDs) exhibit usually soft edges,
\cite{west96,west98} which can be modeled by a harmonic potential confinement.
\cite{kouw01,reim02,yann07} Nevertheless important theoretical studies concerning topological aspects
of {\it nonrelativistic\/} electrons in finite systems under strong magnetic fields have been 
performed by assuming hard-wall boundaries. Well known among such studies are the investigations
\cite{halp82,stre84,stre87,hout89,mont08,mont11} (initiated by Halperin \cite{halp82}) regarding the edge 
states related to the integer-quantum-Hall-effect (IQHE) and those \cite{imry88,imry89,avis93,tana98}
(initiated by Sivan and Imry \cite{imry88}) on the Aharonov-Bohm (AB) 
oscillations which are superimposed on the de Haas - van Alphen 
(dHvA) oscillations. Halperin introduced a hard boundary through an infinite-box-type confining potential, 
while Sivan and Imry used a 10 $\times$ 10 square-lattice tight-binding (TB) model.

Finite graphene nanosystems [Graphene QDs (GQDs) or nanoflakes] offer a broader framework to study
such topological connections. Most importantly, original trends and phenomena can emerge 
\cite{dres96,yann10,roma11} which have no analog with the physics of semiconductor QDs. Indeed, compared to 
SQDs, Graphene QDs exhibit distinct features such as: 1) They possess \cite{yann10,roma11,andr12} atomically
defined sharp physical boundaries (because of the abrupt termination of the honeycomb lattice). 2) Due to 
the underlying honeycomb lattice of graphene, the confined electrons are most appropriately described by TB 
modeling,\cite{dres96,zhan08,roma11,geim09}, while at the same time the corresponding 
continuous description reveals that they behave as massless {\it relativistic\/} particles obeying the 
Dirac-Weyl (DW) equation. \cite{roma11,geim09,aban06,brey06.2} 3) The natural shapes of GQDs are not 
circular, but triangular, hexagonal, or rhombus-like; \cite{roma11} as a result, the electronic 
spectra can explore geometric symmetries lower than the circular one.\cite{roma12} 4) As we will 
show below, the presence of defective edges introduces a quantum-wire and/or ring-type  
({\it doubly-connected-geometry}) behavior, in addition to the {\it singly-connected\/} QD behavior
familiar from the theory\cite{imry89} of SQDs with sharp edges.  

\begin{figure}[t]
\centering\includegraphics[width=6.0cm]{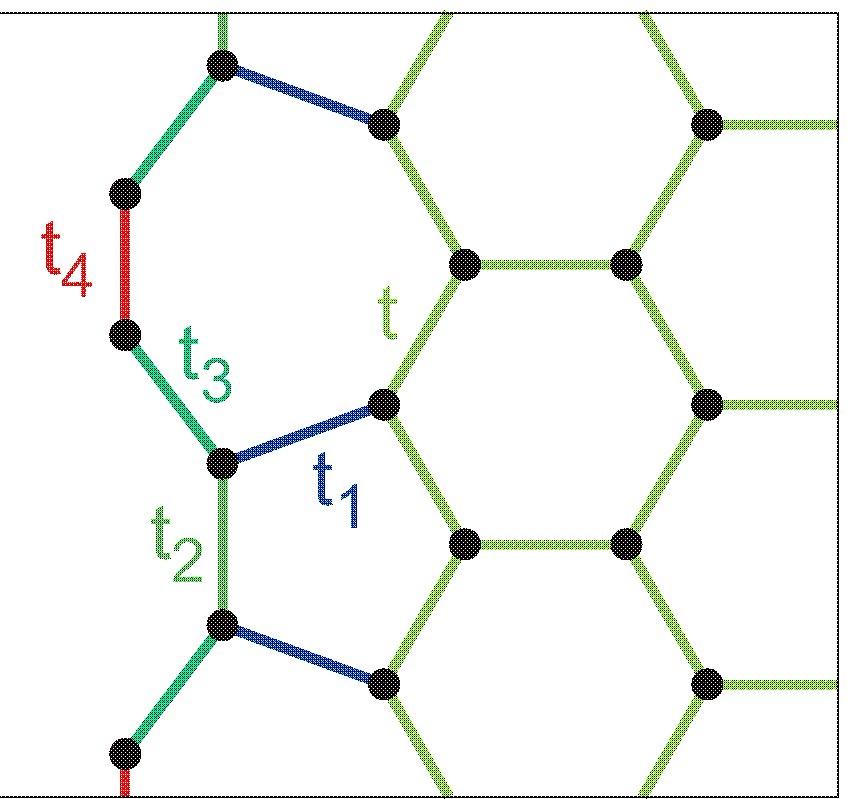}
\caption{
(Color online) Distribution of the hopping matrix elements $t_k$ (see Table \ref{tval}) for the
reczag edge.
}
\label{tis}
\end{figure}

\begin{figure}[t]
\centering\includegraphics[width=8.0cm]{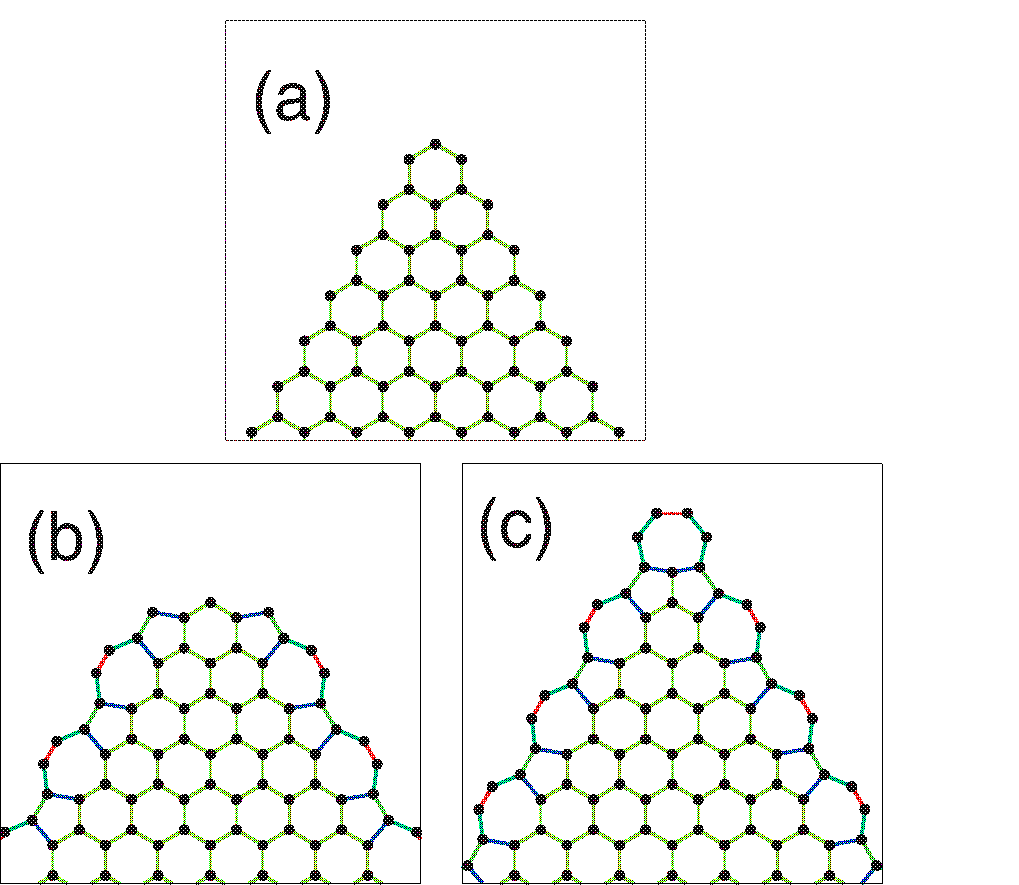}
\caption{
(Color online) Diagrams of corners used for the equilateral trigonal graphene flakes.
(a) corner for zigzag edges. (b) Type-I corner for reczag edges. (c) Type-II corner for reczag edges.
}
\label{corners}
\end{figure}

\subsection{Main findings}
\label{secfind}

The main findings of the paper are:

(I) Beyond the well known features found for graphene quantum dots with zigzag edge terminations, the 
electronic spectra (as a function of the magnetic field $B$) of trigonal graphene nanoflakes with 
reconstructed edges (that is, edge termination with 5-7 defects; see Fig.\ \ref{tis} and Fig.\ \ref{corners}) 
exhibit unique additional regimes; they 
break the particle-hole symmetry and are characterized by a nonlinear dispersion of the electron energy 
versus momentum, associated with a nonrelativistic quantum mechanical description.

\begin{figure*}[t]
\centering\includegraphics[width=14.0cm]{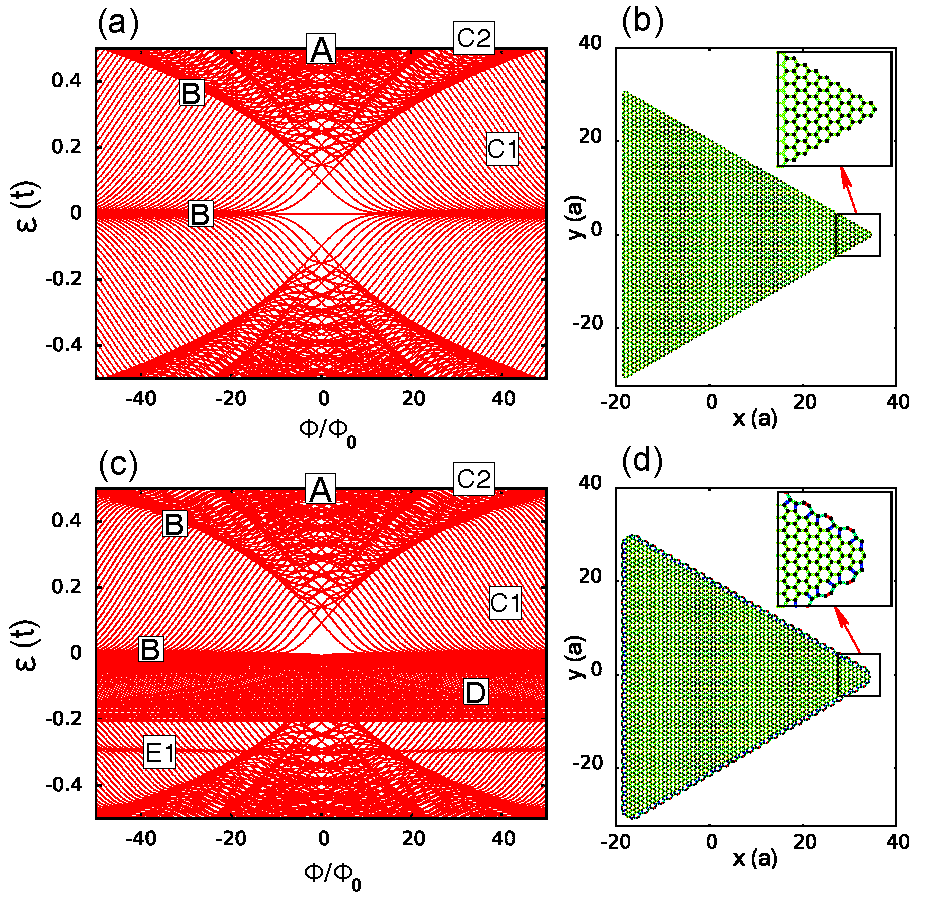}
\caption{
(Color online) 
(a) TB single-particle spectrum for a zigzag trigonal graphene dot as a function of the magnetic field 
(the magnetic flux $\Phi$ over the whole dot). 
(b) Shape of the corresponding equilateral trigonal graphene dot with zigzag edges; it has 61 hexagons in 
the outer row along each side (the total number of carbon atoms is 3966).
(c) TB single-particle spectrum for a type-I reczag trigonal graphene dot as a function of the
magnetic field (the magnetic flux $\Phi$ over the whole dot).
(d) Shape of the corresponding trigonal graphene dot with reczag edges (type-I corner); it has 60 hexagons
in the outer unreconstructed row along each side (the total number of carbon atoms is 4731).
Energy in units of the tight-binding hopping-parameter $t=2.7$ eV. Lengths in units 
of the honeycomb graphene lattice constant $a=0.246$ nm. 
The magnetic flux is given in units of $\Phi_0=hc/e$.
}
\label{trizzrz}
\end{figure*}

(II) The general features shared by graphene flakes with reczag termination with those having zigzag edges,
include the appearance of energy gaps at zero and low magnetic fields due to the finite size (designated as 
region A, see Fig.\ \ref{trizzrz}), the formation of relativistic 
Landau levels (labeled as regions B, see Fig.\ \ref{trizzrz}) at high magnetic fields, and the presence 
between the Landau levels of edge states (socalled Halperin states, labeled as regions $C_i$, see Fig.\ 
\ref{trizzrz}) associated with the IQHE. The characteristic length-scale\cite{mont10} for the Halperin-type 
edge states is the cyclotron radius (magnetic length $l_B$) of the electron orbit (inversely proportional to
the strength of the applied magnetic field). 
 
(III) The unique regimes that emerge in the spectrum of GQDs with reczag (reconstructed zigzag) edges 
include: (a) several features within a band of negative energies 
$ \varepsilon_b=-0.205 t < \varepsilon  < 0$ [region labeled as 
D below in Fig.\ \ref{trizzrz}(c); divisible in regions D1 and D2, see Fig.\ \ref{zo-trirz}], and (b) a 
feature forming a constant-energy line at $\varepsilon_c \approx -0.297t$ [region labeled as E1, see Fig.\ 
\ref{trizzrz}(c)]. The $\varepsilon_b$ lower bound of the region D [see (a) above] is independent of size.
 
(IV) Prominent among the features within the aforementioned $ \varepsilon_b=-0.205 t < \varepsilon < 0$ 
energy stripe is the formation of three-member braid bands (subregion D1, see Fig.\ \ref{zo-trirz}), similar
to those present\cite{baha09,roma12} in the spectra of narrow graphene {\it nanorings\/}, which were shown 
to be associated with Aharonov-Bohm oscillations in graphene nanosystems.\cite{roma12} This suggests that 
the reczag edge behaves in a manner that is analogous to a nanoring enclosing the magnetic flux $\Phi$ 
through the entire area of the graphene flake; $\Phi$ will be given in units of $\Phi_0=hc/e$. 
Obviously the length scale governing the behavior of these edge states associated with the reczag defective
edge is the characteristic length ${\cal L}$ of the entire graphene flake.
This analogy is further substantiated with an analysis using a simple nonrelativistic 1D superlattice model
(see Sec.\ \ref{secqwmod}) where the corners of the trigonal flake are modeled by appropriate scatterers.

(V) Another prominent feature within the $ \varepsilon_b=-0.205 t < \varepsilon < 0$ energy stripe is a 
subregion (D2) of Halperin-type edge states with enhanced density below the zero-Landau level; see
Fig. \ref{zo-trirz} and Sec.\ \ref{secd2}. 

(VI) Furthermore, there are features resulting from localization of the Dirac quasiparticles at the
corners of the polygonal flake (regions labeled as E1 and E2, see Sec.\ \ref{sece1e2}).
 
(VII) A main finding concerns the limited applicability of the continuous Dirac-Weyl equation. As we 
explicitly show in Sec.\ \ref{secdw}, the general features, e.g., the relativistic Landau levels, and the 
Halperin-type edge states, are also present in the continuum-DW reczag spectra. However, concerning the 
unique features found via TB calculations, only the feature of the Halperin-type edge states with an 
enhanced density spectrum (D2 region) maintains also in the continuum spectra; the rest of the special 
reczag features [see (III), (IV), and (VI) above] are missing in the continuum- DW spectrum.
Due to this major discrepancy between the TB and continuum descriptions, we are led to conclude that the 
linearized DW equation fails to capture essential nonlinear physics (i.e., a nonlinear dispersion of
energy versus momentum\cite{note3} coexisting with the Dirac cone), resulting from the introduction of a 
nontrivial (multiple) topological defect\cite{vozm07,bern08,vozm10,mesa09,note4} (e.g., reconstructed reczag
edge) in the honeycomb graphene lattice.

\subsection{Plan of paper}
\label{secplan}

In addition to this section, the Introduction consisted of three other ones: The first (Sec.\ 
\ref{secnano}) provided background information concerning the different graphene edge terminations and 
their nanoelectronics potential, while the second (Sec.\ \ref{sectopo}) introduced the topological aspects. 
The main findings of this paper were outlined in Sec.\ \ref{secfind}.

The remaining of the paper is organized as follows:

Sec.\ \ref{secmeth} recapitulates briefly the tight-binding and continuum Dirac-Weyl methodologies. 

Our main results from the tight-binding calculations are presented in Sec.\ \ref{sectb}. This section is
further divided in two parts: Sec.\ \ref{secgene} describes the general features of the spectra of trigonal
flakes which are shared with GQDs having other edge terminations (e.g., zigzag or armchair). The special
features which are unique to the reczag edge termination are presented in Sec.\ \ref{secspec}. For a
synopsis of these general and special features, see the section describing the main findings (Sec.\ 
\ref{secfind}). Three different sizes of trigonal graphene flakes are considered in Sec.\ \ref{sectb}, with
the two smaller sizes being discussed in Sec.\ \ref{secsmall}.

The corresponding continuous Dirac-Weyl description for a circular reczag GQD is elaborated and contrasted 
to the TB results in Sec.\ \ref{secdw}.

A summary and discussion of our results is given in Sec.\ \ref{secsum}.

Finally the Appendix presents the explicit expressions for the transfer matrices employed in Sec.\ 
\ref{secqwmod}.

\section{Methodology}
\label{secmeth}

In previous publications, we studied primarily graphene quantum dots and graphene nanorings with zigzag 
edge terminations. In this paper, we carry out systematic investigations of the electronic properties of 
graphene flakes with reczag edge terminations and the shape of a regular triangle (see Fig.\ \ref{tis} and
Fig.\ \ref{corners}), and for the cases of zero-magnetic, low-magnetic, and high-magnetic fields. 
In particular, we study the excitation spectra using independent-particle treatments, i.e., we use both 
the tight-binding approach and the semianalytic continuum Dirac-Weyl equations; see, e.g., Refs.\ 
\onlinecite{roma11,roma12}. 

\subsubsection{Basic elements of TB approach}
\label{eltb}

To determine the single-particle spectrum [the energy levels $\varepsilon_i(B)$] in the 
tight-binding calculations for the graphene nanoflakes, we use the hamiltonian
\begin{equation}
H_{\text{TB}}= - \sum_{<i,j>} \tilde{t}_{ij} c^\dagger_i c_j + h.c.,
\label{htb}
\end{equation}
with $< >$ indicating summation over the nearest-neighbor sites $i,j$. The hopping
matrix element 
\begin{equation}
\tilde{t}_{ij}=t_{ij} \exp \left( \frac{ie}{\hbar c}  \int_{{\bf r}_i}^{{\bf r}_j} 
d{\bf s} \cdot {\bf A} ({\bf r}) \right), 
\label{tpei}
\end{equation}
where ${\bf r}_i$ and ${\bf r}_j$ are the positions of the carbon atoms
$i$ and $j$, respectively, and  ${\bf A}$ is the vector potential associated with the
applied perpendicular magnetic field $B$. In the case of a zigzag edge termination, $t_{ij}=t=
2.7$ eV. In the case of the reconstructed reczag edge, four additional values (see Fig.\ \ref{tis})
for the hopping matrix elements must be considered for carbon pairs participating in
the defective edge.\cite{kosk08,osta11} These values are listed in Table \ref{tval}.

\begin{table}[b] 
\caption{\label{tval}%
DFT extracted values for the hopping matrix elements $t_k$ (see Fig.\ \ref{tis}) in the TB 
modeling of a reczag edge, according to Ref.\ \onlinecite{kosk08}. 
}
\begin{ruledtabular}
\begin{tabular}{cccc}
$t_1/t$ & $t_2/t$ & $t_3/t $ & $t_4/t$ \\ \hline 
0.91    & 0.99    & 0.97     & 1.5
\end{tabular}
\end{ruledtabular}
\end{table}

The diagonalization of the TB hamiltonian [Eq.\ (\ref{htb})] is implemented with the use of the 
sparse-matrix solver ARPACK.\cite{arpack} We note here that, unlike the
continuous Dirac-Weyl equations,\cite{rech07,roma11} both the $K$ and $K^\prime$ valleys are 
automatically incorporated in the tight-binding treatment of graphene sheets and nanostructures.

\subsubsection{Basic elements of continuous Dirac-Weyl equations}
\label{eldw}

In polar coordinates, the low-energy noninteracting graphene electrons (around a given $K$ or $K^\prime$ 
point) are most often described via the continuous DW equation.\cite{geim09} Circular symmetry leads to 
conservation of the total pseudospin\cite{geim09} $\hat{\jmath}=\hat{m}+\hat{\sigma}_z$, where $\hat{m}$ is 
the angular momentum and $\hat{\sigma}_z$ the spin of a Dirac electron. 
The reczag edge does not couple the two valleys,\cite{osta11} and
as a result, we seek solutions for the two components $\Psi_A ({\bf r})$ and $\Psi_B ({\bf r})$ or
$\Psi^\prime_A ({\bf r})$ and $\Psi^\prime_B ({\bf r})$ of the single-particle electron orbital (a spinor).
The indices $A$ and $B$ denote the two graphene sublattices and the unprimed and primed symbols are
associated with the $K$ and $K^\prime$ valleys. 

Below we focus on the $K$ valley; similar equations apply also to the $K^\prime$ valley. In polar 
coordinates, one has:
\begin{equation}
\psi_m({\bf r})=
\left( 
\begin{array}{c}
\Psi_A({\bf r})\\
\Psi_B({\bf r})
\end{array}
\right)=
\left(
\begin{array}{c}
e^{im\theta} \chi_A(r)\\
i e^{i(m+1)\theta} \chi_B(r)
\end{array}
\right).
\label{sp}
\end{equation} 
The angular momentum $m$ takes integer values; for
simplicity in Eq.\ (\ref{sp}) and in the following, the subscript $m$ is omitted in 
the sublattice components $\Psi_A$, $\Psi_B$ and $\chi_A$, $\chi_B$.      
 
With Eq.\ (\ref{sp}) and a constant magnetic field $B$ (symmetric gauge), the DW equation 
reduces (for the $K$ valley) to
\begin{eqnarray}
\frac{d}{dx} \chi_B + \frac{1}{x} \left( m+1 + \frac{x^2}{2} \right) \chi_B
&=& \varepsilon \chi^A \nonumber \\
\frac{d}{dx} \chi_A - \frac{1}{x} \left( m + \frac{x^2}{2} \right) \chi_A
&=& - \varepsilon \chi_B; 
\label{dweq}
\end{eqnarray}  
where the reduced radial coordinate $x=r/l_B$ with $l_B=\sqrt{\hbar c / (eB)}$ the
magnetic length. The reduced single-particle eigenenergies $\varepsilon=E/(\hbar v_F/l_B)$, 
with $v_F$ the Fermi velocity. 

The solutions of the DW equations for both valleys in the case of a circular GQD with a reczag edge is
presented in detail in Sec.\ \ref{secdw}.

\section{Tight-binding description for reczag trigonal flakes}
\label{sectb}

\subsection{General features}
\label{secgene}

An example for a trigonal quantum flake is given in Fig.\ \ref{trizzrz} where the single-particle spectrum
(as a function of the magnetic field) of a dot with reczag edges (and type-I corners; see Fig.\ 
\ref{corners}) is compared to that of a dot of similar size, but with unreconstructed zigzag edges. 
Various aspects of trigonal GQDs with pure zigzag edges have been studied earlier; 
\cite{pala07,pota10,ezaw10}
however, for completeness and to allow ready comparisons to be made, we display and briefly comment on the 
corresponding spectrum [see Fig.\ \ref{trizzrz}(a)].
In particular, we have marked main features (or regimes) of the 
zigzag spectrum as follows: The regime of zero and low-magnetic fields is denoted by ``A''; it exhibits 
energy gaps due to finite-size effects. The regime of Landau levels (LLs) formed at high magnetic fields is
denoted by ``B'' (only the $n=0$ and $n=-1$ levels are denoted). The ``C$_i$'s'' denote the edge states
\cite{yann10,roma11} which connect the $|i-1|\text{-}th$ and $|i|\text{-}th$ LLs. The general regimes A, B,
and C are also present in the spectra of trigonal flakes with reczag edges, as an inspection of Fig.\ 
\ref{trizzrz}(c)] readily reveals. 

We note that the three regimes A, B, and C$_i$ have corresponding analogs in the case of a QD with 
nonrelativistic electrons confined by a hard-wall boundary.\cite{imry88,imry89,avis93,tana98} 
These analogies exist despite the well-known differences arising from  the relativistic
nature of Dirac electrons, e.g., the energies of the Landau levels in graphene are 
$E_n= \sgn (n) v_F \sqrt{2e \hbar B |n|}$, $n=0,\pm 1,\pm 2,\ldots$, (square-root $B$-dependence) 
compared to $E_n= \hbar \omega_c (n+1/2)$, $n=0,1,2,\ldots$, [with $\hbar \omega_c=eB/(m^*c)$, 
linear dependence on $B$] for the case of a nonrelativistic 2D electron gas. Such analogs emerge from 
underlying universal and topological properties of the 2D finite systems under high magnetic fields, 
i.e., when $l_B \equiv \sqrt{\hbar c/(eB)}< {\cal L}$ with ${\cal L}$ being a characteristic
length of the nanosystem. Naturally, the energy of the LLs depends on the cyclotron orbit alone, 
and thus it is independent of the size and shape of the dot. But also, this size-and-shape 
independence is shared to a large degree\cite{note1} by the Halperin-type edge states between 
LLs,\cite{halp82} whose energy can be derived (to the lowest order) from a semiclassical or WKB
quantization of a {\it single arc\/} of the skipping orbits, both for nonrelativistic
\cite{hout89,mont08,mont11} and Dirac electrons.\cite{mont10}

\begin{figure}[t]
\centering\includegraphics[width=8.4cm]{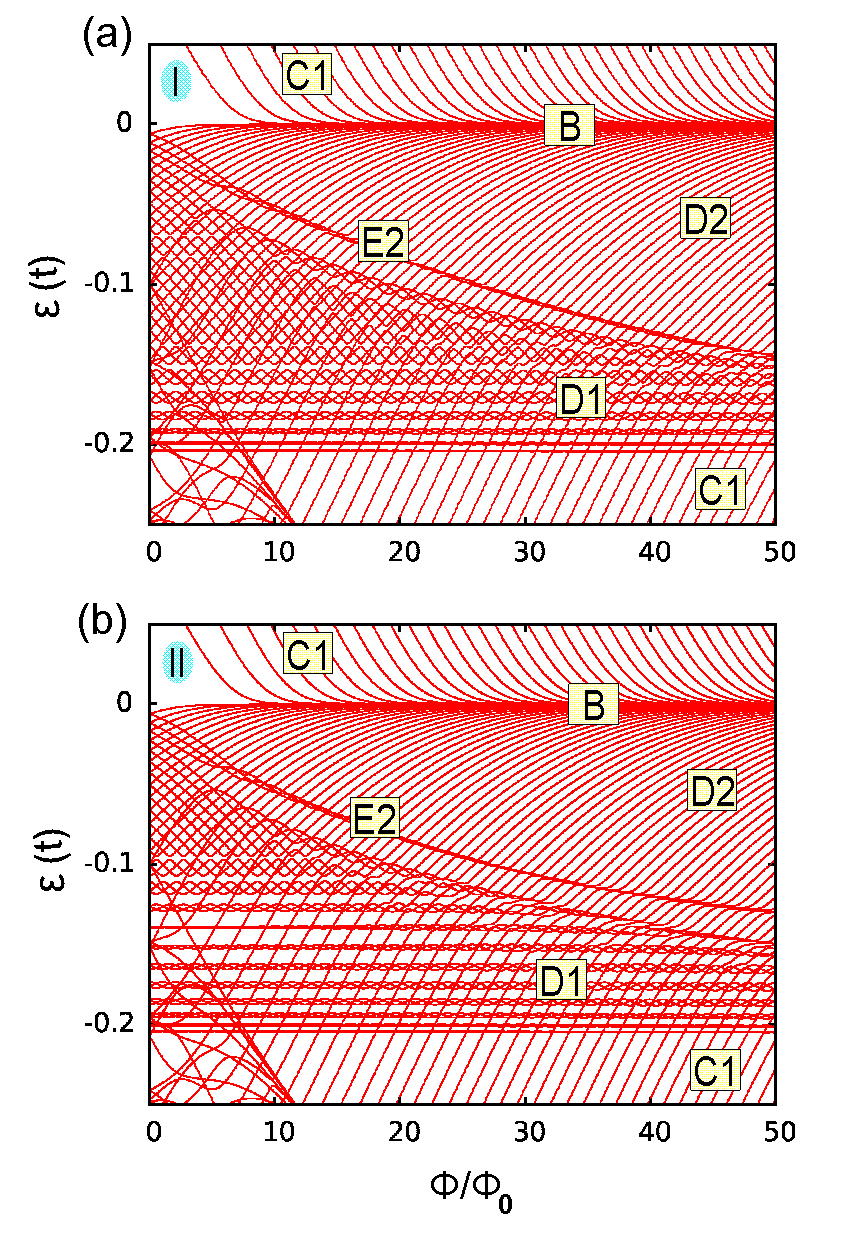}
\caption{
(Color online) 
(a) An enlarged section of the regime marked as D in Fig.\ \ref{trizzrz}(c), showing the TB single-particle 
spectrum for a reczag trigonal graphene dot (with type-I corners), as a function of the magnetic field 
(the magnetic flux $\Phi$ over the whole dot). 
(b) The TB spectrum for the corresponding reczag trigonal GQD with type-II corners.  
Energy in units of the tight-binding hopping-parameter $t$.
The magnetic flux is given in units of $\Phi_0=hc/e$.
}
\label{zo-trirz}
\end{figure}

\begin{figure}[t]
\centering\includegraphics[width=8.0cm]{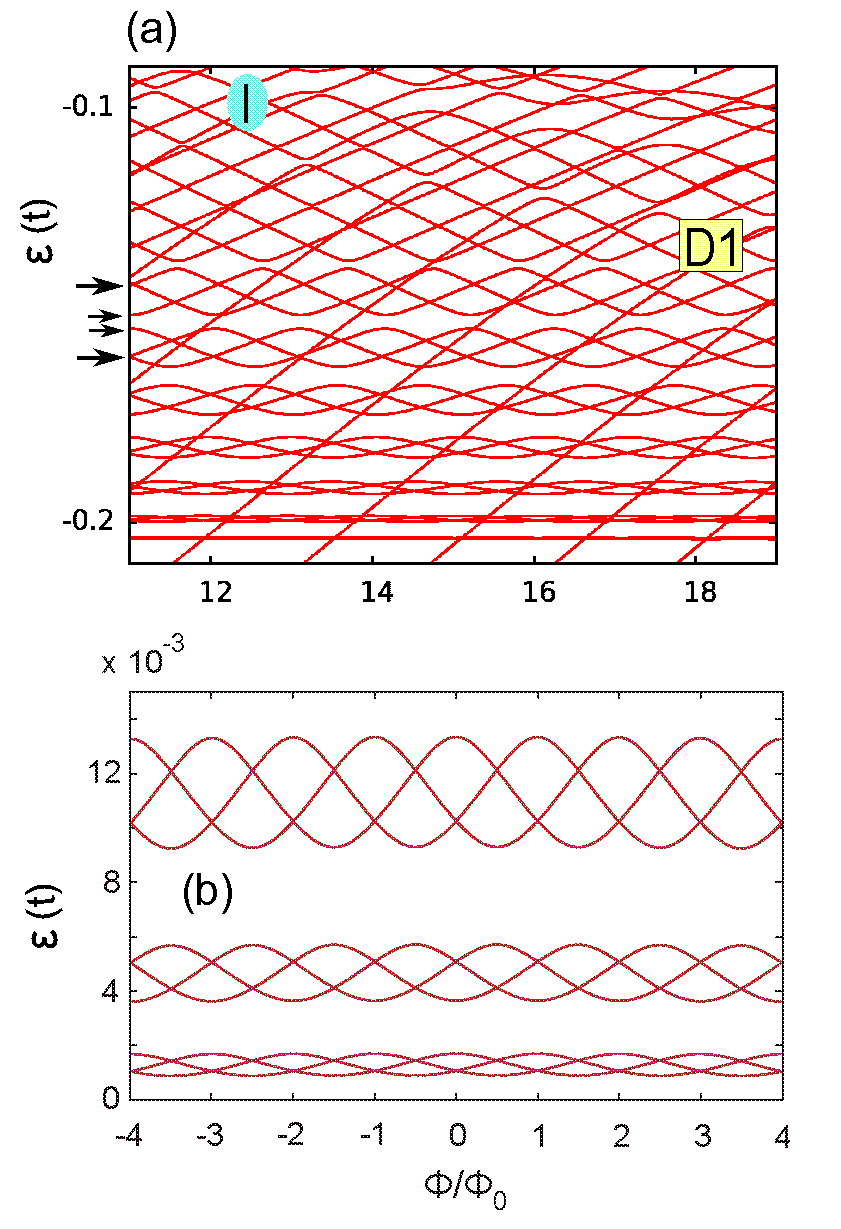}
\caption{
(Color online) 
An enlarged section of the regime marked as D1 in Fig.\ \ref{zo-trirz}(a) showing the TB 
single-particle spectrum for a reczag trigonal graphene dot (with type-I corners), as a function of the 
magnetic field (the magnetic flux $\Phi$ over the whole dot).
The horizontal arrows highlight the alternation 2-1-1-2 (1-2-2-1) in the state degeneracy between two
successive braid bands at $\Phi/\Phi_0=n$ ($\Phi/\Phi_0=n+1/2$), $n=0,1,2,\ldots$.
(b) An example (for reasons of comparison) of a TB single-particle spectrum for a narrow trigonal graphene 
{\it ring\/} with {\it zigzag\/} edge terminations. Such nanorings were used in Ref.\ \onlinecite{roma12} to
study the Aharonov-Bohm oscillatory patterns in graphene nanosystems.
}
\label{zo+trirz-i}
\end{figure}

Of interest for the present study are the Aharonov-Bohm-type refinements concerning the Halperin-type
edge states investigated\cite{imry88,imry89,avis93,tana98} for the case of SQDs. Indeed, Refs.\
\onlinecite{imry88} and \onlinecite{imry89} argued that, in the case of the finite, singly-connected 
QDs, the Halperin-type edge states form an {\it effective\/} ring; in a semiclassical picture they 
correspond to {\it grazing orbits} (see also Ref.\ \onlinecite{blasc97}), reminiscent of the
{\it whispering gallery} trajectories\cite{boga72} investigated at low magnetic fields. As a result the
associated spectra must exhibit a dependence on the total magnetic flux through the area of the QD,
which leads to the emergence of AB-type oscillations in the total Landau magnetization of the dot. 
Specifically these high-$B$ AB oscillations are superimposed on the much larger de Haas - van Alphen
ones, and they tend to decrease as $B$ increases. 

It is apparent, that similar high-$B$ AB-type effects are also present in the case of GQDs with
zigzag and reczag terminations: for example, for the GQDs associated with Figs.\ \ref{trizzrz}(a)
and  \ref{trizzrz}(c), it suffices to calculate the Landau magnetization assuming a zero-temperature 
canonical ensemble and a number, $N$, of Dirac electrons large enough so that the corresponding Fermi level
$\varepsilon_F > 0.2t $.

We stress that our findings regarding trigonal flakes with reczag edges go beyond (see \ref{secspec}) 
the general features described above. Indeed, one of our main findings is that trigonal flakes support,
in addition to the high-$B$, singly-connected-dot AB behavior, oscillatory behavior similar to the
low-$B$ Aharonov-Bohm effect, familiar from semiconductor\cite{imry83,loss91,glaz09,imry09} and
graphene nanorings.\cite{rech07,roma12} The coexistence, in the same nanostructure, of these two distinct 
AB behaviors (associated with singly-connected and doubly-connected geometries) has no analog in
previously considered nanosystems, and it is a special feature unique to graphene defective edges.
        
\subsection{Unique features due to the reczag edge}
\label{secspec}

Having discussed the common general features shared by both the zigzag and reczag trigonal graphene flakes
(see Sect.\ \ref{secgene}), we turn now to study the unique features emerging solely in the case of 
reczag trigonal flakes. An inspection of the electronic spectra in Figs.\ \ref{trizzrz}(a) and 
\ref{trizzrz}(c) shows that the main differences arise from the presence of the two regimes denoted as D 
and E1 in the case of the reczag dot. In particular, the regime D consists of the features within a band of 
negative energies $ \varepsilon_b=-0.205 t < \varepsilon < 0$, while the regime E1 consists of a 
constant-energy line at $\varepsilon_c \approx -0.297t$. (The reconstructed reczag edge violates particle-hole
symmetry, while as is well known the zigzag edge preserves it.) We found that the lower energy bound 
$\varepsilon_b$ of the D regime is independent of the size and shape (e.g., hexagonal versus trigonal 
flake), as well as of the type of corners (type-I versus type-II, see Fig.\ \ref{corners}); 
$\varepsilon_b$ depends only on the values of the TB hopping matrix elements $t_k$ (see Table \ref{tval}).

\begin{figure}[t]
\centering\includegraphics[width=7.0cm]{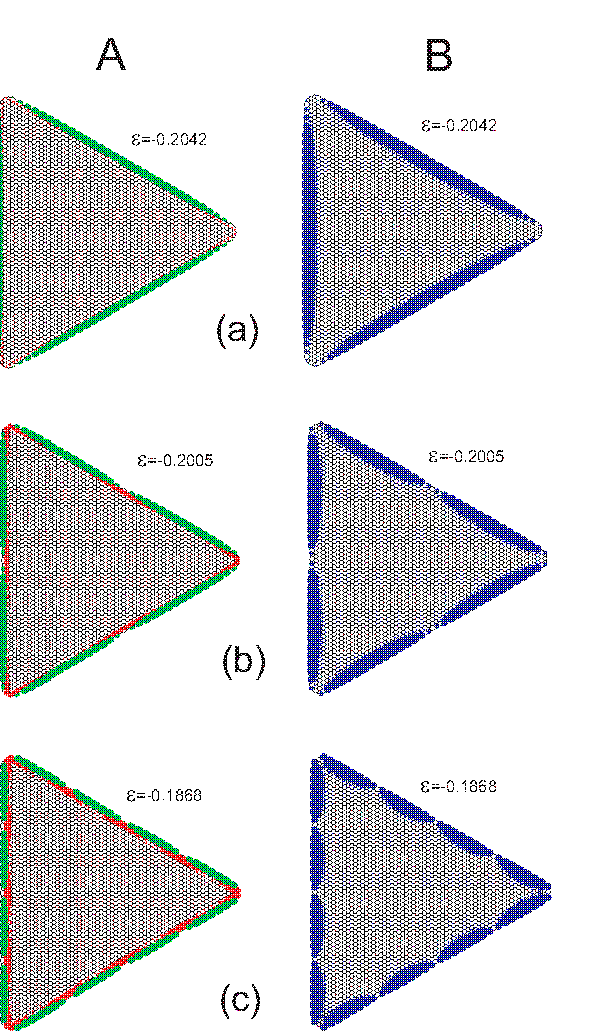}
\caption{
(Color online) 
TB electron densities (modulus) of reczag edge states participating (counting from the bottom) in the first
(a), second (b), and fourth (c) braid bands of region D1 [see Fig.\ \ref{zo+trirz-i}(a)], 
at $\Phi=15.9 \Phi_0$. The A (red) and B (blue) sublattices are plotted separately. Green color denotes the
density on the outer carbon dimers resulting from the edge reconstruction and connected by the hopping 
matrix element $t_4$ in Fig.\ \ref{tis}. (Note that the color codings between Fig.\ \ref{tis} and the 
current Fig.\ \ref{denstrirz} are unrelated). The presence of azimuthal (along the sides of the triangle) 
nodes in the electronic densities is clearly visible. The number of nodes changes by unity from one braid 
band to the next, increasing with increasing energy.  This behavior (including the fact that all three 
states within each braid band maintain the same number of nodes) is quite analogous to that of the edge 
states of a trigonal graphene {\it nanoring\/} at low magnetic fields (see Fig.\ \ref{denstriring} below).
Energies in units of $t=2.7$ eV. 
}
\label{denstrirz}
\end{figure}

\begin{figure}[t]
\centering\includegraphics[width=7.0cm]{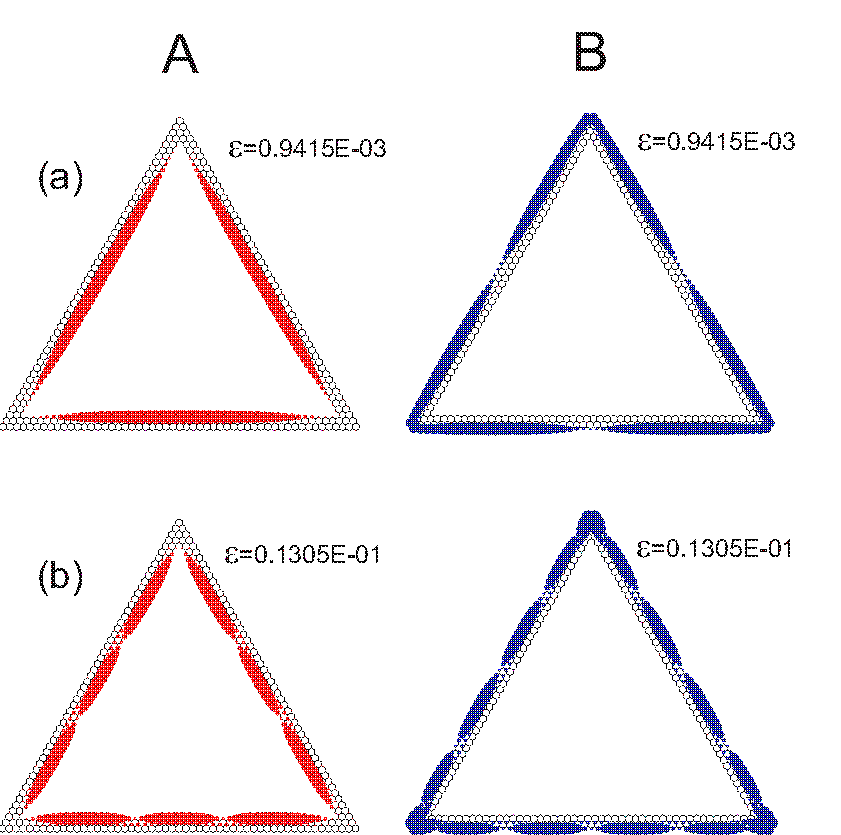}
\caption{
(Color online) 
TB electron densities (modulus) of edge states of the zigzag ring participating (counting from the bottom) 
in (a) the first and (b) the third braid bands in  Fig.\ \ref{zo+trirz-i}(b)], at $\Phi/\Phi_0 \sim 6$. The 
A (red) and B (blue) sublattices are plotted separately. The azimuthal nodes in the electronic densities are
clearly visible. The number of nodes changes by one from one braid band to the next, increasing with 
increasing energy. Energies in units of $t=2.7$ eV. The ring has a total of 906 carbon atoms.
}
\label{denstriring}
\end{figure}

An enlarged section of the electronic spectrum in Fig.\ \ref{trizzrz}(c) (case of type-I corner) is 
displayed in Fig.\ \ref{zo-trirz}(a), while the corresponding section for a trigonal reczag QD with type-II
corners is displayed in Fig.\ \ref{zo-trirz}(b). From a comparison of the two cases in Fig.\ \ref{zo-trirz},
we conclude that the main features in the region D maintain: they show rather small variations between      
the type-I and type-II corners. The larger variation is exhibited by the E1 regime (not shown in
Fig.\ \ref{zo-trirz}). Indeed the E1 line for the type-II corners has moved to a positive energy
$\varepsilon_c \approx 0.120t$. The enlarged spectra in Fig.\ \ref{zo-trirz} suggest a further division of 
the D regime into features denoted as D1, D2, and E2. (The grouping of the E2 feature with the E1 feature 
will become apparent below; see Sec. \ref{sece1e2}).
   
Because of the similarity between the electronic spectra of the two types of corners, it will be sufficient 
below to restrict our further analysis of spectral features to the case of type-I corners [see Fig.\ 
\ref{zo-trirz}(a) and Fig.\ \ref{trizzrz}(c)].

\subsubsection{Region D1: Ideal-ring, low-$B$-type edge states and Aharonov-Bohm oscillations} 
\label{secd1}

The main feature of the D1 region are the many energy bands consisting of three-curve braid patterns, an 
enlargement of which is displayed in Fig.\ \ref{zo+trirz-i}(a). These braid bands are quite similar to the 
ones displayed by the low-$B$ electronic spectra of a narrow trigonal graphene {\it nanoring\/} with zigzag
edges [see Fig.\ \ref{zo+trirz-i}(b)], which were investigated\cite{roma12} 
recently in the context of the AB effect. Based on this 
similarity and the findings of Ref.\ \onlinecite{roma12}, we infer that these braid bands are associated 
with the formation of a second type of edge states, in addition to the Halperin-type ones. This second type
edge states are localized (in the radial direction) within the physical defective reczag edge and exhibit 
behavior associated with a quantum wire. In particular, in the case of a trigonal reczag-GQD, the three wire 
segments along the sides of the triangle are coupled pairwise (via electron tunneling at the corners) and 
form a trigonal quantum nanoring. Henceforth, we will adopt the term reczag edge states to designate these 
states, which are associated with the physical defective edge.   

To gain further insight into the similarity of the reczag edge states to the graphene-ring states, we 
display in Fig.\ \ref{denstrirz} the probability densities at $\Phi/\Phi_0=15.9$ ($\phi/\Phi_0=0.0037$) for 
several of the reczag states [with energies belonging to successive braid bands starting with the 
lowest-in-energy one; see Fig.\ \ref{zo+trirz-i}(a)]; $\phi$ denotes the magnetic flux through a single 
hexagon of the honeycomb graphene lattice. Probability densities at $\Phi/\Phi_0 \sim 6$
($\phi/\Phi_0=0.003$) for two characteristic states of the narrow trigonal graphene nanoring with zigzag 
edges (considered in Ref.\ \onlinecite{roma12}) are displayed in Fig.\ \ref{denstriring}. It is apparent 
that the electronic densities in Fig.\ \ref{denstrirz} (reczag flake) are restricted near the physical
boundary of the flake, and thus they correspond to formation of edge states. In addition, the presence of 
azimuthal (along the sides of the triangle) nodes in these electronic densities is clearly visible, and the 
number of nodes changes by unity from one braid band to the next, increasing with increasing energy. This 
behavior (including the fact that all three states within each braid band maintain the same number of 
azimuthal nodes) is quite analogous to that of the edge states of a trigonal graphene {\it nanoring\/} at 
low magnetic fields (see Fig.\ \ref{denstriring}). 

The similarities between the reczag edge states and the low-$B$ states of graphene nanorings indicates 
that the reczag edge behaves like a quantum wire. Naturally, this quantum-wire behavior places the reczag 
edge states in a separate category, different from that of the Halperin-type edge states. In Sec. 
\ref{secqwmod} below, we will further elaborate on the quantum-wire aspects of the reczag edge states 
using a simple one-dimensional superlattice model.     

\begin{figure}[t]
\centering\includegraphics[width=7.0cm]{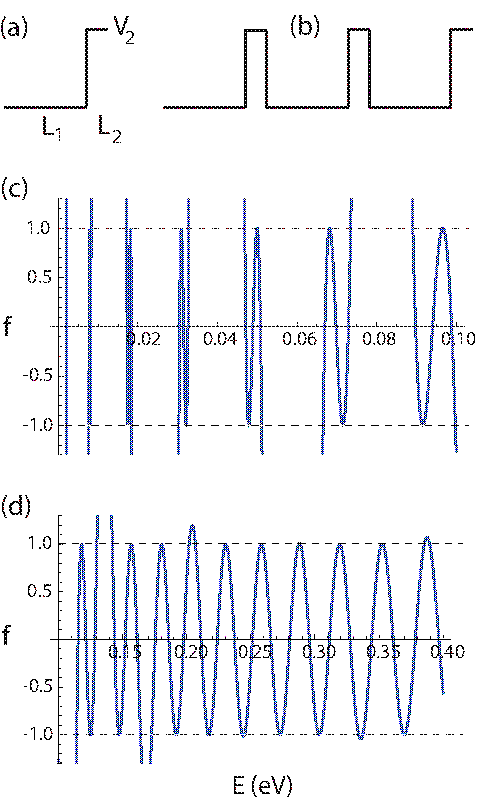}
\caption{
(Color online) 
(a) Schematic diagram of the unit subcell associated with a single side of the trigonal graphene flake 
with reczag edge terminations. $V_2$ denotes the height of the potential barrier which mimicks the 
scatterer at the corners of the trigonal reczag flake. 
(b) Schematic diagram of the unit cell associated with the magnetic-field virtual lattice; it involves 
all three sides of the equilateral triangle, and thus it consists of three unit subcells in a series. 
(c)-(d) The function $f(E)=\Tr[{\bf T}(E)]/2$ [see the r.h.s. of Eq.\ (\ref{disrel})], which is associated 
with the unit cell [shown in (b)] of the magnetic-field virtual superlattice,
as a function of the energy variable $E$. In calculating $f(E)$, the parameters 
for the unit subcell [shown in (a)] were taken as: $L_1=12.5$ nm, $L_2=1.5$ nm, $V_1=0$, and $V_2=0.1$ eV. 
The relevant values of $f(E)$ lie within $\pm 1$ (i.e., within the dotted lines). (c) $f(E)$ in the range 
$0 \leq E \leq V_2$. (d) $f(E)$ in the range $V_2  \leq E \leq 4V_2$. Energy in units of eV.
}
\label{super}
\end{figure}

\begin{figure}[t]
\centering\includegraphics[width=7.0cm]{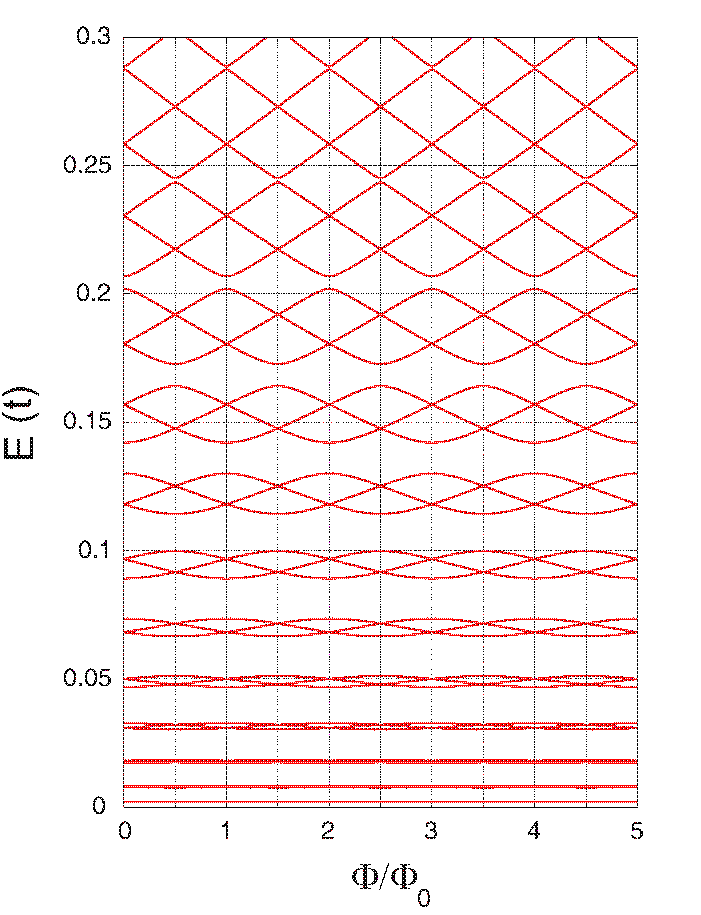}
\caption{
(Color online) 
Single-particle spectrum (as a function of the total flux $\Phi$) from the semianalytic  superlattice 
model considered in Sec.\ \ref{secqwmod}. The parameters for the unit subcell mimicking each reczag side of
the trigonal graphene flake are: $L_1=12.5$ nm, $L_2=1.5$ nm, $V_1=0$, and $V_2=0.1$ eV. Note that the 
total length $L_1+L_2=14$ nm is similar to the length of the trigonal flake in Fig.\ \ref{trizzrz}. The 
height of the potential barrier defining the scatterers at the corners $(V_2)$ is roughly one fifth 
of the width ($0.2t$) of the D region [see Fig.\ \ref{trizzrz}(c)]. Energies in units of eV.
The magnetic flux is given in units of $\Phi_0=hc/e$.
}
\label{qwmod}
\end{figure}

\subsubsection{A simple semianalytic model for the reczag edge states}
\label{secqwmod}

In this section we show that the main qualitative features of the braid bands in the D1 region can be 
reproduced using a simple nonrelativistic 1D superlattice approach. Indeed, in this approach, each side of 
the trigonal reczag flake is modeled as a unit subcell consisting of a two-region piece-wise potential [see 
Fig.\ \ref{super}(a)]. In particular, the first and wider region was chosen to have a length of $L_1=12.5$ 
nm and a zero potential height, $V_1=0$. The second region models the scatterer's behavior of the 
triangle's corner and was taken to be a narrow potential barrier; we chose $L_2=1.5$ nm and $V_2=0.1$ eV. 
Note that the total length $L=L_1+L_2=14$ nm is similar to the length of the side of the equilateral 
triangle in Fig.\ \ref{trizzrz}, while the height of the potential barrier is roughly one fifth of the 
width ($0.2t$) of the D region (see Fig.\ \ref{trizzrz}). Naturally, due to the simplicity of the model, we
did not attempt to achieve a full quantitative agreement with the TB spectra.  

Following Ref.\ \onlinecite{gilmbook}, one constructs first the transfer matrices ${\bf M}_1$ and 
${\bf M}_2$ (see the Appendix) for the regions 1 and 2 of the unit subcell portrayed in Fig.\ 
\ref{super}(a).  Then the transfer matrix (${\bf T_s}$) for the unit subcell is simply the product of the 
two matrices ${\bf M}_1$ and ${\bf M}_2$, i.e.,
\begin{equation}
{\bf T_s} = {\bf M}_1 {\bf M}_2.
\label{ts}
\end{equation}

A magnetic field perpendicular to the plane generates a flux $\Phi$ over the entire area of the flake. Thus
all three sides of the triangle must be considered in the study of magnetic-field effects. To this end, 
and following Ref.\ \onlinecite{imry83}, we consider the equivalent problem of a magnetic-field virtual 
superlattice. In our case, however, the unit cell of the virtual lattice is more complex; it consists of 
three unit subcells in a series [see Fig.\ \ref{super}(b)] in order to account for the three scatterers at 
the corners. Then the transfer matrix for the unit cell is given by
\begin{equation}
{\bf T} = {\bf T_s}^3.
\label{tt}
\end{equation}

To form the magnetic-field superlattice, the unit cell must be repeated ad-infinitum. This is equivalent to
imposing periodic boundary conditions on a succession of finite lattice blocks with ${\cal N}$ unit cells
and taking the limit ${\cal N} \rightarrow \infty$. Accordingly,\cite{gilmbook} the dispersion relation 
determining the energy bands of the virtual superlattice is given by
\begin{equation}
\cos(2 \pi \Phi/\Phi_0) = \Tr[{\bf T}(E)]/2,
\label{disrel}
\end{equation}
where we used the fact that the equivalent Bloch wave vector for the magnetic-field superlattice is 
${\cal K}=2 \pi \Phi/(3L \Phi_0)$, $3L$ being the width of the unit cell (see Ref.\ \onlinecite{imry83}). 

The energy bands resulting from the dispersion relation in Eq.\ (\ref{disrel}), with the specific parameter
values mentioned in the beginning of this section, are displayed in Fig.\ \ref{qwmod}. A comparison with
the braid bands in Figs.\ \ref{zo-trirz}(a) and \ref{zo+trirz-i}(a) (D1 region of the TB spectra) shows 
that the simple 1D model reproduces the essential trends of the TB braid bands. Specifically,
the common trends are as follows: (I) The alternation 2-1-1-2 (1-2-2-1) in the state degeneracy between two 
successive braid bands at $\Phi/\Phi_0=n$ ($\Phi/\Phi_0=n+1/2$), $n=0,1,2,\ldots$ [see the horizontal 
arrows in Fig.\ \ref{zo+trirz-i}(a) and Fig.\ \ref{qwmod}]. (II) The width of the braid bands increases 
with increasing energy. (III) In contrast, the energy gaps separating the braid bands decrease with 
increasing energy. (IV) At high enough energies, the braid bands tend to merge into a single pattern having
``chicken-wire'' topology, familiar from the well-known ideal-metal-ring energy spectrum;\cite{cheu88} this
last feature is present in the TB spectra of Fig.\ \ref{zo-trirz} in the region $0 \leq \Phi/\Phi_0 < 8.0$. 

We note that in the context of the simple 1D model of this section, these trends can be further understood 
from an inspection of the behavior of the $f(E)$ function plotted in Fig.\ \ref{super}(c) and Fig.\ 
\ref{super}(d). Indeed, for a given $\Phi$, the single-particle energies plotted in Fig.\ \ref{qwmod} 
correspond to the crossing points of the $f(E)$ curve with a horizontal straight line having an ordinate 
$f=\cos(2 \pi \Phi/\Phi_0) < 1$. In particular, the trend No. IV above is associated with the asymptotic 
behavior of the $f(E)$ function; this asymptotic behavior at high energies (above the barrier height $V_2$)
corresponds to the fact that the tunneling particle behaves like a free fermion and it does not feel 
strongly the effect of the scatterers. 

\begin{figure}[t]
\centering\includegraphics[width=7.0cm]{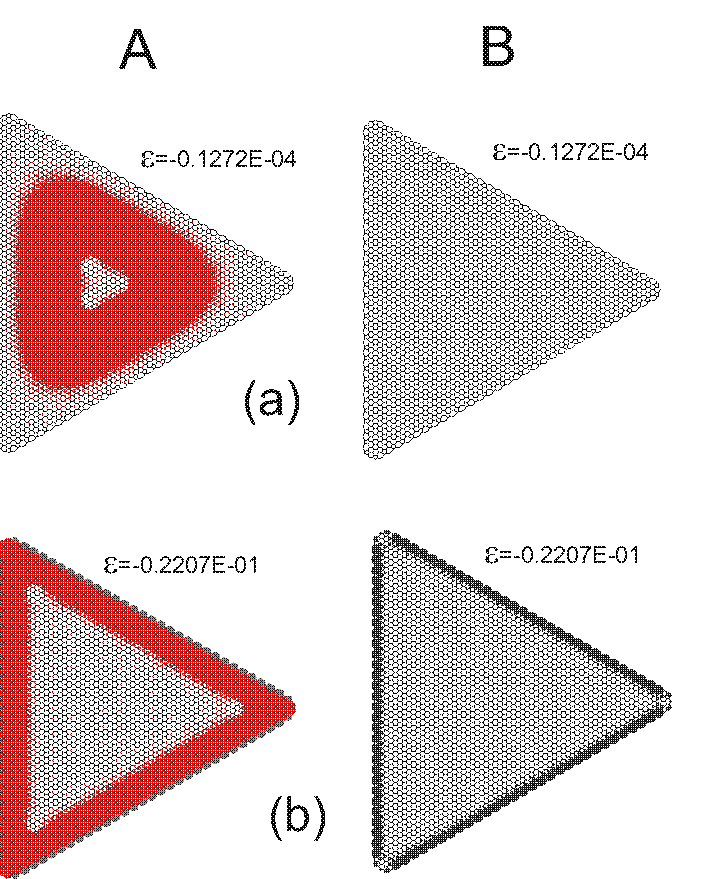}
\caption{
(Color online) 
TB electron densities (modulus) of two characteristic states for the D2 regime of the reczag flake at 
$\Phi/\Phi_0=15.9$.
(a) a state with near-zero energy $\varepsilon=-0.1272 \times 10^{-4}t$ exhibiting bulk zero-Landau-level 
behavior.  
(b) a state with lower energy $\varepsilon=-0.2207  \times 10^{-1}t$ exhibiting Halperin-edge behavior.
Compared to Fig.\ \ref{denstrirz}, the absence of azimuthal nodes in the electronic densities here is 
noticeable. Energies in units of $t=2.7$ eV. The total number of carbon atoms is 4731. 
The A (red) and B (blue) sublattices are plotted separately.
Green color denotes the density on the outer carbon dimers resulting from the edge reconstruction and 
connected by the hopping matrix element $t_4$ in Fig.\ \ref{tis}.
}
\label{denstrirzd2}
\end{figure}

\begin{figure}[t]
\centering\includegraphics[width=7.0cm]{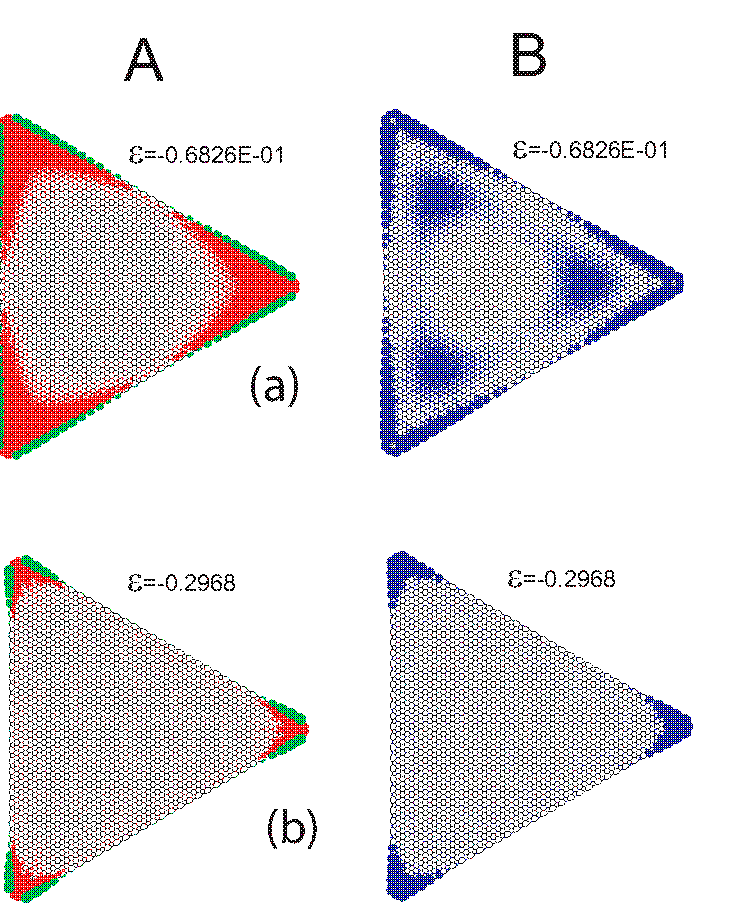}
\caption{
(Color online) 
TB electron densities (modulus) of two characteristic states for the E1 and E2 regimes of the reczag flake 
at $\Phi/\Phi_0=15.9$. Note the concentration of the electron densities at the corners of the triangle.
(a) a state in the E2 regime with energy  $\varepsilon=-0.6826 \times 10^{-1}t$. 
(b) a state in the E1 regime with energy $\varepsilon = -0.297t$.
Energies in units of $t=2.7$ eV. The A (red) and B (blue) sublattices are plotted separately.
Green color denotes the density on the outer carbon dimers resulting from 
the edge reconstruction and connected by the hopping matrix element $t_4$ in Fig.\ \ref{tis}.
}
\label{denstrirze1e2}
\end{figure}

\begin{figure*}[t]
\centering\includegraphics[width=14.0cm]{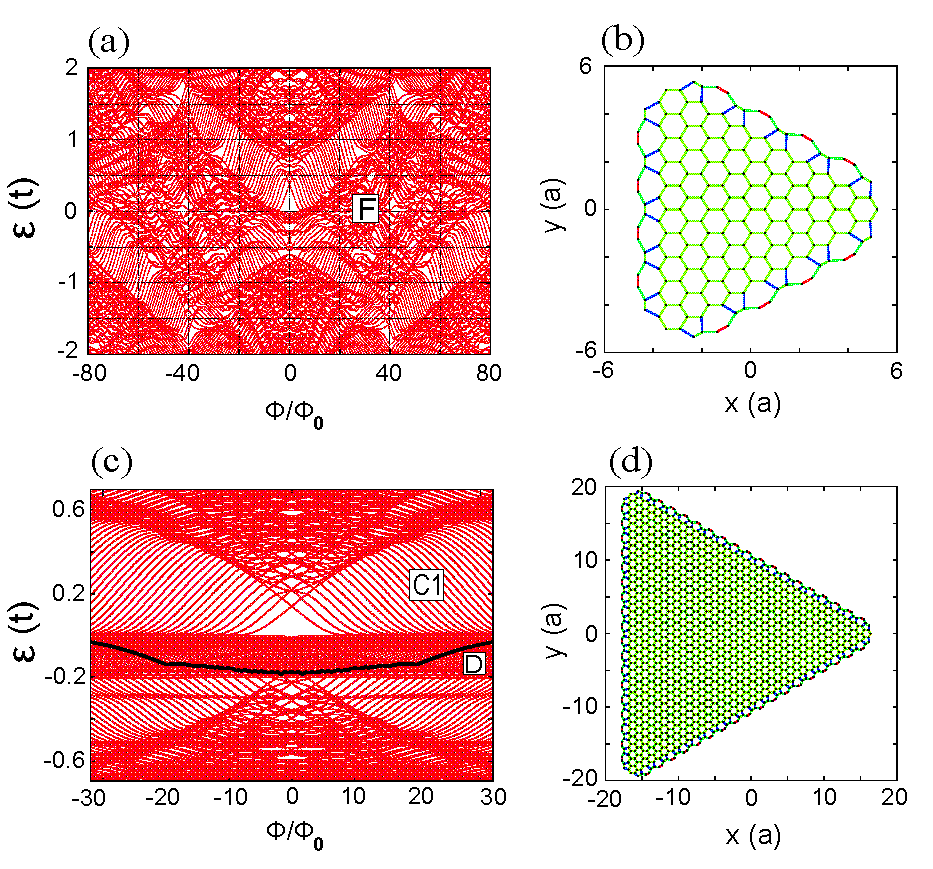}
\caption{
(Color online) 
(a) TB single-particle spectrum for a very small (with type-I corners) reczag trigonal graphene dot, as a 
function of the magnetic field (the magnetic flux $\Phi$ over the whole dot). 
(b) Shape of the equilateral trigonal graphene dot which corresponds to (a); it has 10 hexagons in the
outer unreconstructed row along each side (the total number of carbon atoms is 195).
(c) TB single-particle spectrum for a larger (with type-I corners) reczag trigonal graphene dot, 
as a function of the magnetic field (the magnetic flux $\Phi$ over the whole dot). 
(d) Shape of the trigonal graphene dot which corresponds to (c); it has 38 hexagons in the outer
unreconstructed row along each side (the total number of carbon atoms is 1819).
The thick black line in (c) denotes the Fermi level in the canonical ensemble corresponding to $N=60$ holes
(spin included).
Energy in units of the tight-binding hopping-parameter $t=2.7$ eV. Lengths in units of the honeycomb
graphene lattice constant $a=0.246$ nm.
The magnetic flux is given in units of $\Phi_0=hc/e$.
}
\label{trirzsm}
\end{figure*}

\begin{figure*}[t]
\centering\includegraphics[width=14.0cm]{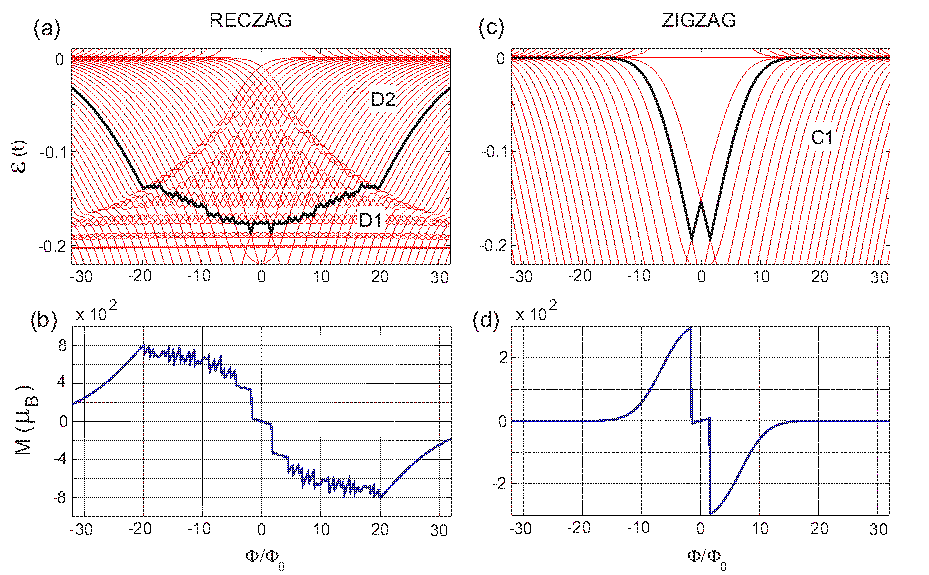}
\caption{
(Color online) 
(a) {\it Reczag flake:\/} An enlarged part of the TB single-particle spectrum shown in Fig.\ \ref{trirzsm}(c),
as a function of the magnetic field (the magnetic flux $\Phi$ over the whole dot). The shape of the 
corresponding reczag flake is displayed in Fig.\ \ref{trirzsm}(d). 
(b) {\it Reczag flake:\/} Landau magnetization (at zero temperature) 
for $N=60$ holes (spin included) exhibiting 
Aharonov-Bohm oscillations superimposed on larger ones generated by the rapid variation of the background 
Halperin-type edge states which cross the braid bands. The thick black line in (a) denotes the corresponding 
Fermi level in the canonical ensemble.
(c) {\it Zigzag flake:\/} A similar part of the TB single-particle spectrum for the corresponding zigzag 
trigonal flake (with 38 hexagons in the outer row along each side), as a function of the magnetic flux $\Phi$.
(d) {\it Zigzag flake:\/} Landau magnetization for $N^*=4$ effective holes (spin included); 
the absence of Aharonov-Bohm 
oscillations is apparent. The thick black line in (c) denotes the corresponding Fermi level in the canonical 
ensemble. For a meaningful comparison, this Fermi level was chosen to fall within the 
$ \varepsilon_b=-0.205 t < \varepsilon  < 0$ energy band. Note that for the reczag flake this energy band 
contains the special D region; for the zigzag flake this energy band is reduced to being part of region C1. 
The total number of holes is $N=N_0+N^*$, where $N_0$ is the number of strictly zero-energy states present in 
the zigzag trigonal flake (also, see text). 
Energy in units of the tight-binding hopping-parameter $t=2.7$ eV.
The magnetic flux is given in units of $\Phi_0=hc/e$.
}
\label{trirzab}
\end{figure*}

\subsubsection{Region D2: Dense spectrum of Halperin-type edge states}
\label{secd2}

We focus now on the region marked as D2 in Fig.\ \ref{zo-trirz}(a). The single-particle spectra in this region
consist of energy curves similar to those of the Halperin-type edge states in region C1 
(which connect the $n=1$ and $n=0$ graphene Landau levels). 
A main difference, however, between these two regions is that the 
spectrum in D2 is more dense compared to that in region C1. For example, at $\Phi/\Phi_0=15.9$, we found
that within the range $|\varepsilon| \leq 0.4414 t$, there are 20 states in the D2 region, but only 10
states in the C1 region above the zero-energy line. We note that the density of states in the C1-region of 
a reczag flake is similar to that in the C1-region of a zigzag flake with comparable size. As a result,
because all the states in region D2 converge to the zero-energy Landau level, the degeneracy 
(density of states per unit magnetic flux) of this Landau level is higher in the case of a trigonal reczag 
flake compared to that of a pure zigzag flake. This behavior raises naturally the question of whether the 
conductance properties of the anomalous\cite{geim09,aban06,brey06.2} relativistic IQHE will be 
impacted. We will, however, defer elaborating on this 
question until the section on the continuous Dirac-Weyl description (Sec. \ref{secdw}).

To further investigate the properties of region D2, we display (for $\Phi/\Phi_0=15.9$) in Fig.\ 
\ref{denstrirzd2} electron densities for a couple of characteristic states in this region.
Compared to Fig.\ \ref{denstrirz}, the absence of azimuthal nodes in their electronic densities is
noticeable. Specifically,, in Fig.\ \ref{denstrirzd2}(a) we consider a state with near-zero energy 
($\varepsilon=-0.1272 \times 10^{-4}t$). This state exhibits a zero-Landau-level behavior familiar from a 
graphene sheet,\cite{geim09} and accordingly, one sublattice component (here the B-sublattice) vanishes 
everywhere. This contrasts with the special case of the zero-Landau-level states in a zigzag flake, which 
are of a mixed bulk-edge character, with the bulk and edge components residing on different sublattices.
\cite{yann10,roma11} In Fig.\ \ref{denstrirzd2}(b), we consider a state with lower energy 
$\varepsilon=-0.2207  \times 10^{-1}t$, which is representative of the pristine Halperin-type double-edge 
states between the $n=0$ and $n=1$ Landau levels discussed in Refs.\ \onlinecite{yann10,roma11} for
GQDs with zigzag edge terminations. 

The enhanced density of TB states in the D2 region maintains also in the spectra derived from the
continuous Dirac-Weyl equation in the case of a circular disk with reczag edges (see Sec.\ \ref{secdw} below).

\subsubsection{Regions E1 and E2: States localized at the corners} 
\label{sece1e2}

The states belonging to the E1 and E2 regimes are grouped together. Indeed, as revealed from the electron
densities displayed in Fig.\ \ref{denstrirze1e2}, they are localized (to one degree or the other) at the
corners of the triangle. As seen from Fig.\ \ref{zo-trirz}(a),  the E2 feature consists of three states
whose energy curves form a single braid, similar to the braids in region D1. One of the states in this
triad (with energy $\varepsilon=-0.6826 \times 10^{-1}t$ at $\Phi/\Phi_0=15.9$) is plotted in Fig.\ 
\ref{denstrirze1e2}(a). Because of the localization at the corners, the quantum-wire model of Sec.\ 
\ref{secqwmod} is not appropriate for the E2 regime. However, as discussed in Sec.\ IV A of Ref.\ 
\onlinecite{yann03} (see in particular Figs.\ 6 and 7 therein), a simple H\"{u}ckel model involving three 
localized Gaussian wave functions at the corners of an equilateral triangle is able to reproduce 
qualitatively the braiding behavior of the energy curves as a function of the magnetic field.  

The states in the E1 regime behave in a different way; in fact, their energies as a function of $B$ 
do not form a braid, but an approximate straight line located at $\varepsilon_c \approx -0.297 t$. In the
C2 region (between the $n=-2$ and $n=-1$ Landau levels), there are three such states with very close 
energies [at $\Phi/\Phi_0=15.9$, these energies are $-0.2957t$, $-0.2968t$, and $-0.2976t$; the state 
corresponding to the second energy here is plotted in Fig.\ \ref{denstrirze1e2}(b)]. In the C1 region
(between the $n=-1$ and $n=0$ Landau levels), only two of these states exist. At present, we are unaware
of any simple model describing such a behavior.

Because the corners were shown earlier to act as scatterers (see Sec.\ \ref{secqwmod}),   
the appearance of states that are localized at (or attracted towards) the corners may seem counterintuitive
at a first glance. This behavior, however, originate from the relativistic nature of the graphene massless
Dirac quasiparticles for which the scatterers may also act as centers of attraction due to Klein 
tunneling.\cite{klei29,kats06} In this context, we mention Ref.\ \onlinecite{kim10}, where similar 
localized wave functions under the repulsive potential barrier defining a circular graphene antidot were 
reported.  

\subsection{Smaller trigonal shapes and Aharonov-Bohm oscillations} 
\label{secsmall}

Of interest is the question of the size-dependence of the spectra of the reczag trigonal flakes. The size 
of the flake investigated in previous sections [with sixty hexagons along each  side, see Fig.\ 
\ref{trizzrz}(d)] is sufficiently large for the main features of the spectra to have been fully developed. 
We thus briefly investigate here smaller sizes. Indeed, Fig.\ \ref{trirzsm}(a) displays the spectrum of a 
very small trigonal reczag flake with 10 hexagons along each side [see the corresponding shape in Fig.\ 
\ref{trirzsm}(b)], while Fig.\ \ref{trirzsm}(c) displays the spectrum of an intermediate-size flake with 38
hexagons along each side [see the associated shape in Fig.\ \ref{trirzsm}(d)].

The spectrum for the very small flake [Fig.\ \ref{trirzsm}(a)] exhibits rather large differences from that
of the large flake [Fig.\ \ref{trizzrz}(c)]. This is mainly due to the full development (within the plotted 
$\Phi$ range) of the Hofstadter-butterfly\cite{hofs76,zhan08} fractal patterns (designated as region F), 
which appear for very strong magnetic fields such that $l_B \lesssim a$, i.e., when the magnetic length is 
similar to or smaller than the honeycomb graphene-lattice constant. 
Furthermore the Landau levels (region B) and
region D (which is unique to the reczag edges and has been our main focus in this paper) are hardly
recognizable; they are strongly quenched compared to the case of the large flake in Fig.\ \ref{trizzrz}(c).

For the intermediate-size case shown in Fig.\ \ref{trirzsm}(c), both the Landau-level regime and the two 
regimes D1 (three-member braid bands) and D2 (Halperin-type edge states with enhanced density) are well 
developed; see enlarged part in Fig.\ \ref{trirzab}(a). We note again the constancy and size-independence 
of the lower bound $\varepsilon_b$ of the D region.

We take advantage of the full development of the spectrum in the intermediate size, and we calculate 
explicitly for this size the Landau magnetization [displayed in Fig.\ \ref{trirzab}(b)] for a positively 
charged flake with $N=60$ holes (spin included). 
Following Ref.\ \onlinecite{roma12}, we carry out this calculation in the 
canonical ensemble and zero temperature, and the thick black line in Fig.\ \ref{trirzab}(a) denotes the 
corresponding Fermi level.
As a function of the total magnetic flux $\Phi$, the magnetization exhibits clear (albeit with variable 
shapes) oscillatory Aharonov-Bohm patterns associated with the braid bands. At the same time, these AB
patterns are superimposed on larger oscillations generated by the rapid variation (with $\Phi$) of 
background Halperin-type edge states crossing the braid bands. These background Halperin edge states are 
also responsible for the skipping of the Fermi level between different braid bands and between different 
states in the same braid band, which results in the jumps and in the variation of the shape of the AB 
patterns (which is to be contrasted with the regular AB oscillations in graphene nanorings with zigzag edges
\cite{roma12}).

We display also in Fig.\ \ref{trirzab}(c) and Fig.\ \ref{trirzab}(d) the energy spectrum and Landau 
magnetization, respectively, for the corresponding zigzag trigonal flake (with 38 hexagons in the outer row 
along each side). The absence of Aharonov-Bohm oscillations in Fig.\ \ref{trirzab}(d) is apparent. For a 
meaningful comparison, the Fermi level in the canonical ensemble [see thick black line in (c)] was chosen to 
fall within the $ \varepsilon_b=-0.205 t < \varepsilon  < 0$ energy band. We note that for the reczag 
flake this energy band contains the special D region; for the zigzag flake this energy band is reduced to 
being part of region C1. For the zigzag flake the Fermi level is determined by the number $N^*$ of effective 
holes ($N^*=4$ here, spin included). Indeed the total number of holes is $N=N_0+N^*$, with $N_0$ being the 
number of strictly zero-energy states present in the zigzag trigonal flake ($N_0$ equals\cite{pala07,pota10} 
the number of hexagons along one side minus one).
Naturally, the strictly zero-energy states do not contribute to the Landau magnetization. We further note that
as a result of the reconstruction process (reczag flake), however, the strictly zero-energy states acquire 
finite energies. In a continuum model (see Sec.\ \ref{secdw} below), this mapping is codified by the boundary 
condition specified by Eq.\ (\ref{conrec}), which involves the reczag parameter ${\cal F}$ [Eq.\ 
(\ref{valf})]; for ${\cal F}=0$, the zigzag-edge case is recovered.   

\begin{figure}[t]
\centering\includegraphics[width=8.0cm]{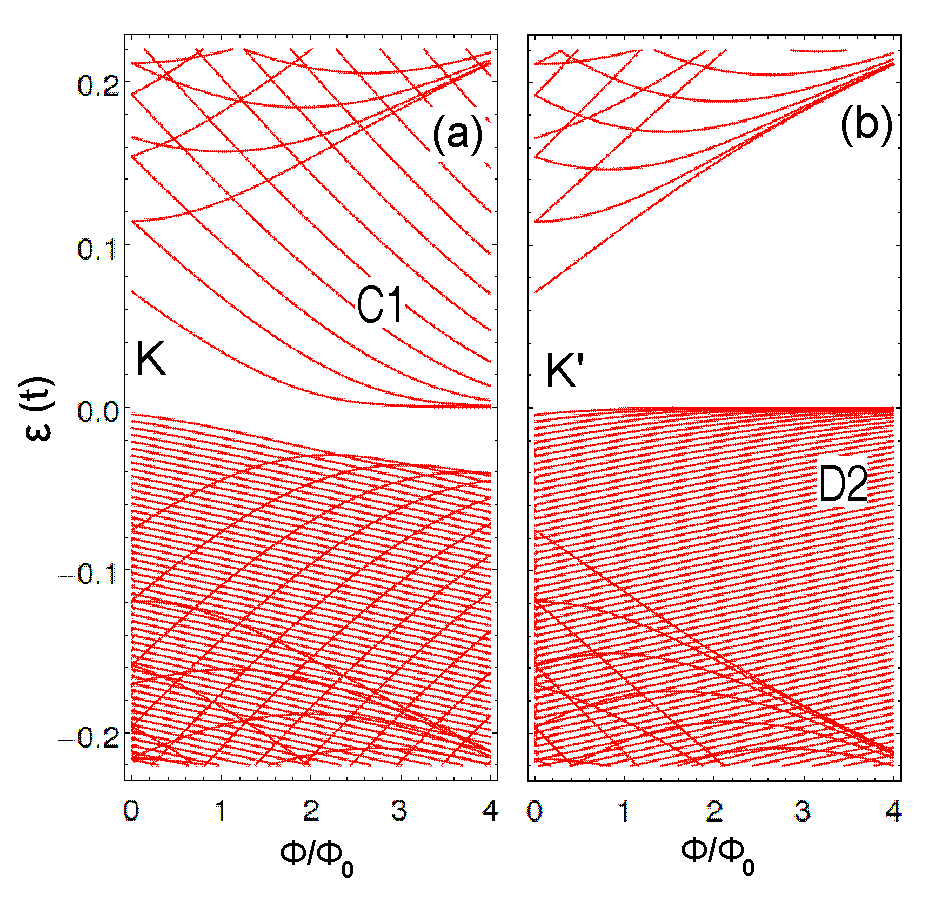}
\caption{
(Color online) 
Single-particle spectra as a function of the magnetic flux $\Phi$, according to the continuous Dirac-Weyl 
description for a circular GQD with a reczag edge termination. (a) The $K$ valley. (b) The $K^\prime$ valley. 
The radius of the dot is $R=8$ nm. Energies in units of $t=2.7$ eV.
}
\label{cirdw}
\end{figure}

\section{Continuous Dirac-Weyl description for circular reczag GQDs}
\label{secdw}

In order to describe the properties of graphene and graphene nanosystems near the neutral Dirac point, 
the continuous Dirac-Weyl equation has been widely and successfully used as an alternative to the TB
calculations. In particular for graphene nanoribbons with zigzag and armchair edge
terminations there is an overall agreement between the TB results and those of the DW approach.
Although the shape of a GQD in the continuous description is most often taken as circular and not 
polygonal, this overall agreement (albeit with certain caveats) between circular and TB calculations was 
also found to extend to the case of graphene nanoflakes and nanodots (see, e.g., Ref. \onlinecite{roma11}).
It is thus of interest to investigate whether such overall agreement applies also for the unique features 
of a reczag flake discussed in earlier sections.

In the continuum approach, a graphene Dirac electron (or hole) is represented by a four-component spinor
$(\Psi_A, \Psi_B, \Psi^\prime_A, \Psi^\prime_B)^T$, with the indices A and B denoting the two 
sublattices, and the unprimed and primed symbols denoting the $K$ and $K^\prime$ valleys.
In the case of zigzag or armchair edge terminations, the four components of the spinor obey well-known 
characteristic boundary conditions.\cite{mont10,geim09,brey06} For the case of the reczag edge, 
corresponding boundary conditions were proposed recently in Ref.\ \onlinecite{osta11}. For the $K$ valley 
these conditions relate the components on the A and B sublattices as follows:
\begin{equation}
\Psi_A = i {\cal F} \Psi_B,
\label{conrec}
\end{equation}
where the parameter ${\cal F}$ is defined as
\begin{equation}
{\cal F} = \frac{t_1^2t_4(t_2t_4-t_3^2)}{2t(t_3^4+t_2t_3^2t_4+t_2^2t_4^2)}.
\label{valf}
\end{equation}
~~\\
The value for ${\cal F}=0.07$ for the reczag edge; see Table \ref{tval} for the values
of the hopping matrix elements $t_k$, $k=1,2,3,4$. 
For the $K^\prime$ valley, the boundary condition is obtained via the substitution ${\cal F} \rightarrow
-1/{\cal F}$. Note that the reczag edge does not mix the two valleys,\cite{osta11} as is the case with the 
zigzag boundary condition.

For a finite circular graphene sample of radius $R$, we seek solutions of Eq.\ (\ref{dweq}) 
for $\varepsilon \neq 0$ that are {\it regular at the origin\/} ($x=0$). For a nanodot with a reczag edge
one finds that the single-particle spectrum is given by the solutions of the following dispersion relation:
\begin{equation}
\chi_B(\varepsilon,m,x) + {\cal F} \chi_A(\varepsilon,m,x)=0,
\label{disrel2}
\end{equation}
where ${\cal F}=0.07$ for the $K$ valley and ${\cal F}=-1/0.07$ for the $K^\prime$ valley, $x=R/l_B$, $m$ is
an angular momentum, and (see Ref.\ \onlinecite{yann10})
\begin{widetext}
\begin{equation}
\chi_A (\varepsilon,m,x) \propto \left\{ 
\begin{array}{cc}
x^m e^{-x^2/4} M( m+1-\frac{\varepsilon^2}{2},m+1,\frac{x^2}{2}); & \;\text{if}\;\; m \geq 0 \\
x^{-m} e^{-x^2/4} M(1-\frac{\varepsilon^2}{2},-m+1,\frac{x^2}{2}); & \;\text{if}\;\; m \leq -1
\end{array}
\right.
\label{chi1}
\end{equation}
and
\begin{equation}
\chi_B (\varepsilon,m,x) \propto \left\{ 
\begin{array}{cc}
\frac{\varepsilon}{2} \frac{x^{m+1}}{m+1} e^{-x^2/4} M( m+1-\frac{\varepsilon^2}{2},m+2,\frac{x^2}{2}); 
& \;\text{if}\;\; m \geq 0 \\
\frac{2m}{\varepsilon}x^{-m-1} e^{-x^2/4} M(-\frac{\varepsilon^2}{2},-m,\frac{x^2}{2}); 
& \;\text{if}\;\; m \leq -1
\end{array}
\right.,
\label{chi2}
\end{equation}
\end{widetext}
where $M(a,b,z)$ is Kummer's confluent hypergeometric function. \cite{abrabook}

\begin{figure}[t]
\centering\includegraphics[width=8.0cm]{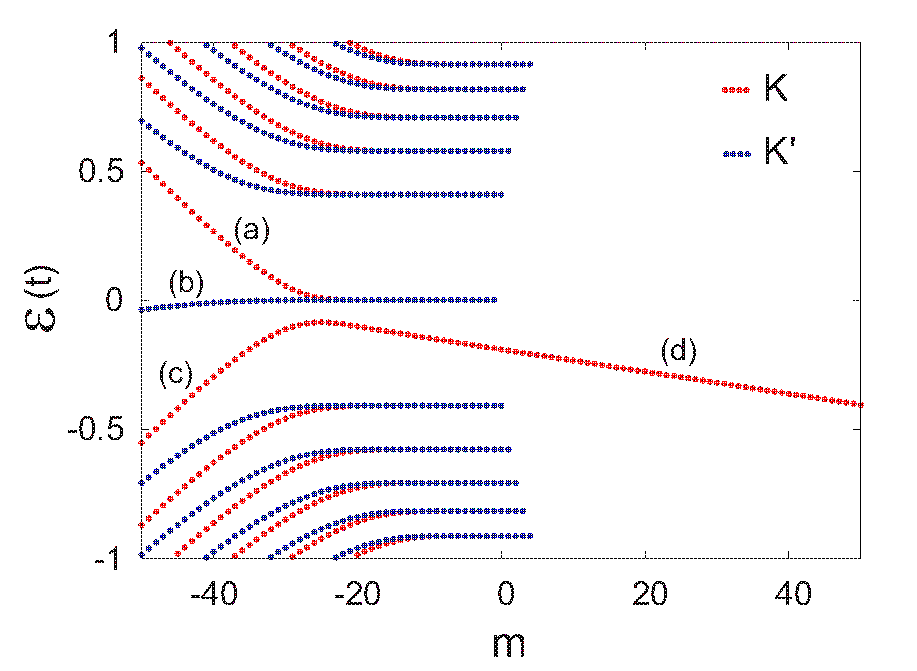}
\caption{
(Color online) 
Single-particle energies as a function of the angular momentum $m$, according to the continuous Dirac-Weyl 
description for a circular GQD with a reczag edge termination. The total flux is fixed at $\Phi=15 \Phi_0$. 
Both the $K$ and  $K^\prime$ valleys are considered. Note the four dispersive branches of edge states [(a), 
(b), (c), and (d)] associated with the zeroth Landau level. We note that only two dispersive branches of
edge states [(a) and (c)], associated with the zeroth Landau level, appear in a circular GQD with a 
{\it zigzag} edge termination. The radius of the dot is $R=8$ nm. Energies in units of $t=2.7$ eV.
}
\label{cirdw2}
\end{figure}

The solutions of the dispersion relation in Eq.\ (\ref{disrel2}) are plotted in Fig.\ \ref{cirdw}(a) for
the $K$ valley and in Fig.\ \ref{cirdw}(b) for the $K^\prime$ valley. One observes that the general
features discussed in Sec.\ \ref{secgene} (namely, the Landau levels, and the Halperin-type edge states) are
also present in the continuum-DW reczag spectra. However, concerning the unique features found via 
TB calculations [Sec.\ \ref{secspec}] and associated with a trigonal reczag flake, only the feature of the 
Halperin-type edge states with an enhanced density spectrum (D2 region) maintains also in the continuum 
spectra [see Fig.\ \ref{cirdw}(b)]. The rest of the special reczag features are missing in Fig.\ 
\ref{cirdw}: in particular we note the nonexistence of a lower-energy bound $\varepsilon_b$ for the D region
and the absence of the three-member braid bands (region D1), the latter being a reflection of the ability of
the defective reczag edge to behave as a 1D quantum nanoring. Furthermore, we note that the E1 and E2
states, which are localized at the corners, are also missing in the continuum model.

Due to these major discrepancies between the TB and continuum descriptions, we are led to conclude that
the linearized DW equation fails to capture essential nonlinear physics resulting from the introduction
of a nontrivial defect in the honeycomb graphene lattice. Indeed the Dirac-Weyl equation is 
obtained for the low-energy states of electrons in the honeycomb lattice, and it is not valid at the reczag 
edges and the corners, where the topological structures are very different from the honeycomb lattice.

As mentioned earlier in Sec.\ \ref{secd2}, the presence of Halperin-type edge states with an enhanced density
spectrum (D2 region) raises naturally the question whether this feature may impact the conductance behavior 
of the anomalous\cite{geim09,aban06,brey06.2} relativistic IQHE. To be able to answer this question within
the continuous DW description, one needs to count the dispersive branches of edge states present in the 
spectrum of the circular reczag dot when the single-particle 
energies are plotted versus the angular momentum $m$ and at a fixed value
of the magnetic flux (the magnetic field). For a circular reczag GQD with radius $R=8$ nm (as was the case
in Fig.\ \ref{cirdw} where the magnetic flux was varied), this latter spectrum is displayed in Fig.\ 
\ref{cirdw2} (for a fixed magnetic flux $\Phi=15 \Phi_0$). Both the $K$ and  $K^\prime$ valleys are 
considered. We note that there are four dispersive branches [labeled as (a), (b), (c), and (d)] associated 
with the zeroth Landau level. Furthermore, it was found that all four channels represent edge states; see
also Ref.\ \onlinecite{osta11} where the case of the linear reczag edge of a semi-infinite graphene plane was
considered. In contrast, only two dispersive branches [corresponding to (a) and (c)], associated with the 
zeroth Landau level, appear in a circular GQD with a {\it zigzag} edge termination.\cite{aban06,brey06.2} The
appearance of these four branches in the spectrum of the circular reczag GQD, however, does not influence the
IQHE conductance, because two of them, i.e., the (c) and (d) are counter propagating, and thus their 
contributions are expected to cancel each other. 

We stress, however, that the above conclusion is based on the continuous DW spectrum. As noted above, the
DW spectrum differs drastically from the TB one, and thus a definitive answer to the question concerning
the IQHE-conductance behavior associated with a trigonal reczag flake requires a full study of the
current/transmission using the tight-binding method.\cite{roma13}   

\section{Summary and discussion}
\label{secsum}
The electronic spectra of graphene nanoflakes with reczag edges, where a succession of pentagons and 
heptagons, that is 5-7 topological defects, replaces the hexagons at the familiar zigzag edge, were 
investigated via systematic tight-binding calculations. Three different sizes of trigonal graphene flakes 
were considered in Sec.\ \ref{sectb}, with the two smaller sizes being discussed in Sec.\ \ref{secsmall}.
(A detailed recapitulation of the results was given in Sec.\ \ref{secfind} of the Introduction.) 
Emphasis was placed on topological aspects and connections underlying the patterns dominating these spectra.
A central result is that the spectra of trigonal reczag flakes exhibit both general features (Sec.\ 
\ref{secgene}), which are shared with GQDs having other edge terminations (i.e., zigzag or armchair),
as well as special ones (Sec.\ \ref{secspec}), which are unique to the reczag edge termination. 
These unique features include breaking of the particle-hole symmetry, and they are associated with a 
nonlinear dispersion of the energy as a function of momentum, which may be interpreted as nonrelativistic 
behavior. 

The general topological features (Sec.\ \ref{secgene}) shared with the zigzag flakes include the appearance 
of energy gaps at zero and low magnetic fields due to finite size, the formation of relativistic Landau 
levels at high magnetic fields, and the presence between the Landau levels of Halperin-type edge states 
associated with the integer quantum Hall effect. Topological regimes, unique to the reczag nanoflakes
(Sec.\ \ref{secspec}), appear within a stripe of negative energies $\varepsilon_b=-0.205t < \varepsilon 
< 0$, and along a separate feature forming a constant-energy line outside this stripe. 

Prominent among the patterns within the $ \varepsilon_b=-0.205t  < \varepsilon < 0$ energy stripe is the 
formation of three-member braid bands, resembling those in the spectra of narrow graphene {\it nanorings\/}
(Sec.\ \ref{secd1}). The reczag edges along the three 
sides of the triangle act as a one-dimenional quantum wire (with the
corners behaving as scatterers) enclosing the magnetic flux through the entire area of the graphene flake 
(Sec.\ \ref{secqwmod}). This leads to the development of Aharonov-Bohm-type oscillations in the magnetization
(Sec.\ \ref{secsmall}). Another prominent feature within the $ \varepsilon_b=-0.205t < \varepsilon < 0$ 
energy stripe is a subregion of Halperin-type edge states of enhanced density immediately below the 
zero-Landau level (Sec.\ \ref{secd2}). Furthermore, there are features resulting from localization of the 
Dirac quasiparticles at the corners of the polygonal flake (Sec.\ \ref{sece1e2}).

A main finding concerns the limited applicability of the continuous Dirac-Weyl equation in conjuntion with 
the boundary condition proposed in Ref.\ \onlinecite{osta11}. Indeed, this combination does not reproduce 
the special reczag features. Due to this discrepancy between the tight-binding and continuum descriptions, 
one is led to the conclusion that the linearized Dirac-Weyl equation fails to capture essential nonlinear 
physics resulting from the introduction of a multiple topological defect in the honeycomb graphene lattice.

We comment here that simpler topological defects (e.g., a single\cite{bern08} pentagon, heptagon, or
pentagon-heptagon pair embedded in the honeycomb lattice) are often described\cite{mesa09,vozm10} 
(at zero magnetic field) in the continuum DW approach via a gauge field (an additional vector potential) 
resembling the one generated by an Aharonov-Bohm magnetic-flux solenoid. The generalization of this 
gauge-field modification of the DW equation to multiple topological defects may provide a better overall 
agreement with the TB results.

\begin{acknowledgments}
This work was supported by the Office of Basic Energy Sciences of the US D.O.E.
under contract FG05-86ER45234. 
\end{acknowledgments}

\appendix*

\section{Expressions for the transfer matrices}

For the first region of the unit subcell in Fig.\ \ref{super}(a), the transfer matrix is \cite{gilmbook}
\begin{equation}
{\bf M}_1 = \left( \begin{array}{cc}
\cos(k_1L_1) & -\sin(k_1L_1)/k_1 \\
k_1 \sin(k_1L_1) & \cos(k_1L_1) \end{array} \right),
\label{mtrx1}
\end{equation}
with $k_1=\sqrt{2mE/\hbar^2}$; $m$ is the nonrelativistic electron mass and $E$ the energy 
variable.

For the second region of the unit subcell, the transfer matrix is\cite{gilmbook}
\begin{equation}
{\bf M}_2 = \left( \begin{array}{cc}
\cosh(\kappa_2 L_2) & -\sinh(\kappa_2 L_2)/\kappa_2 \\
-\kappa_2 \sinh(\kappa_2 L_2) & \cosh(\kappa_2 L_2) \end{array} \right),
\label{mtrx21}
\end{equation}
with $\kappa_2=\sqrt{2m(V_2-E)/\hbar^2}$, if $E < V_2$, and
\begin{equation}
{\bf M}_2 = \left( \begin{array}{cc}
\cos(k_2 L_2) & -\sin(k_2 L_2)/k_2 \\
k_2 \sin(k_2 L_2) & \cos(k_2 L_2) \end{array} \right),
\label{mtrx22}
\end{equation}
with $k_2=\sqrt{2m(E-V_2)/\hbar^2}$, if $E \geq V_2$.
 
Using the matrices ${\bf M}_1$ and ${\bf M}_2$ defined above, and with the help of the algebraic language 
MATHEMATICA,\cite{math} we found that the trace of the transfer matrix ${\bf T}$ 
[see Eqs.\ (\ref{ts}) $-$ (\ref{disrel})], which is associated with the unit cell of the virtual magnetic 
superlattice, is given by 
\begin{widetext}
\begin{eqnarray}
\Tr[{\bf T}(E)] &=& \left \{ 2 k_1^3 \kappa_2^3 \cos(3 k_1 L_1) \cosh^3(\kappa_2 L_2) + 
  3 k_1^2 \kappa_2^2 (-k_1^2 + \kappa_2^2) \cosh^2(\kappa_2 L_2) \sin(3 k_1 L_1) \sinh(\kappa_2 L_2) + 
\right. \nonumber \\ 
&&  3 k_1 \kappa_2 \cos(k_1 L_1) \biglb( k_1^4 + \kappa_2^4 - 
  (k_1^2 - \kappa_2^2)^2 \cos(2 k_1 L_1) \bigrb) \cosh(\kappa_2 L_2) \sinh^2(\kappa_2 L_2) + 
\nonumber \\ 
&&  \left. (k_1^2 - \kappa_2^2) \biglb(-3 (k_1^2 + \kappa_2^2)^2 \sin(k_1 L_1) 
  + (k_1^2 - \kappa_2^2)^2 \sin(3 k_1 L_1)\bigrb) \sinh^3(\kappa_2 L_2)/4 \right \}/(k_1^3 \kappa_2^3),
\end{eqnarray}
when $E < V_2$, and

\begin{eqnarray}
\Tr[{\bf T}(E)] &=& 
- \left \{-2 k_1^3 k_2^3 \cos^3(k_1 L_1) \cos(3 k_2 L_2) + \right. \nonumber\\
&& \left. 3 k_1 k_2 \cos(k_1 L_1) \cos(k_2 L_2) 
\biglb(-k_1^4 - k_2^4 + (k_1^2 + k_2^2)^2 \cos(2 k_2 L_2)\bigrb) \sin^2(k_1 L_1) + \right. \nonumber \\ 
&&  \sin^3(k_1 L_1) \sin(k_2 L_2) 
\biglb(-3 k_1^2 k_2^2 (k_1^2 + k_2^2) \cos^2(k_2 L_2) + (k_1^6 + k_2^6) \sin^2(k_2 L_2)\bigrb) +
\nonumber \\ 
&& \left. 3 k_1^2 k_2^2 (k_1^2 + k_2^2) \cos^2(k_1 L_1) \sin(k_1 L_1) \sin(3 k_2 L_2) \right \}/(k_1^3 k_2^3),
\end{eqnarray}
when $E \geq V_2$.
\end{widetext}


\begin{thebibliography}{999}
\bibitem{note3}
Strictly speaking, the energy versus linear momentum curve, $E(k)$, corresponds to the case of a straight
reczag edge in a semi-infinite graphene plane, and was calculated in Ref.\ \onlinecite{osta11}; see Fig.
3(b) therein. However, Ref.\ \onlinecite{osta11} passed over the importance of the nonlinear behavior of
this curve. For a reczag graphene flake, the effects associated with this nonlinear behavior become
paramount.
\bibitem{note4}
Using tight-binding, Ref.\ \onlinecite{bern08} studied single topological defects (one pentagon, one 
heptagon, and a pentagon-heptagon pair) embedded in a graphene nanoribbon. The reczag edge studied here is 
a multiple topological defect.
\bibitem{note1} This size-and-shape independence for the Halperin-type edge states becomes absolute
in the limit $B \rightarrow \infty$ (or $l_B/{\cal L} \rightarrow 0$), when the curvature of the boundary
can be neglected.
\end{thebibliography}
\end{document}